\documentclass[aps,physrev,reprint,groupedaddress]{revtex4-2}
\usepackage{amsmath,graphicx}
\usepackage{subcaption}
\usepackage{placeins}
\usepackage[usenames,dvipsnames]{color}
\graphicspath{{figures_blank/}}
\begin{document}

\title{Nonlinear Localized States on a Pyrochlore Lattice}

\author{Faustino Palmero Ramos}
\affiliation{Department of Mathematics and Statistics, University of Massachusetts Amherst, Amherst, 01003-4515, MA, USA}

\author{Avadh Saxena}
\affiliation{Center for Nonlinear Studies and Theoretical Division, Los Alamos National Laboratory, Los Alamos, New Mexico 87545, USA}

\author{Panayotis G. Kevrekidis}
\affiliation{Department of Mathematics and Statistics, University of Massachusetts Amherst, Amherst, 01003-4515, MA, USA}
\affiliation{Department of Physics, University of Massachusetts Amherst, Amherst, 01003-4515, MA, USA}

\date{\today}

\begin{abstract}
In the present work we explore a prototypical three-dimensional (3d) lattice
possessing a flat band in the form of a pyrochlore lattice in the context 
of a dispersive nonlinear dynamical model, namely the discrete nonlinear
Schr{\"o}dinger (DNLS) equation. We set up the corresponding steady state and dynamical
problems and discuss the linear spectrum of the relevant model before delving
into a more detailed analysis of the nonlinear equilibria of the system. For the latter,
we analyze the more well-established ---at the DNLS level--- fundamental discrete
soliton states, as well as vortex structures. For the fundamental solitary waves,
we connect their existence and stability with how they approach the linear bands.
In the vortex case,  we identify their stability features for vortices
of topological charge $S=1$ and $S=2$ with those of the 
honeycomb and triangular lattices. An arguably even more intriguing feature of
the pyrochlore lattice concerns the compactly supported
nonlinear eigenstates stemming from the flat band of the linear spectrum.
These compact localized modes are found to possess oscillatory instabilities
for a range of propagation constants in the focusing case, although they can
be stable in the latter, while they are found to be
subject to symmetry-breaking instabilities in the
defocusing nonlinearity case. These results offer a glimpse at the nexus 
of topology, flat band systems and dispersive nonlinear lattices in three spatial
dimensions and as such may be a starting point toward a deeper exploration
of such an intriguing interplay.
\end{abstract}

\maketitle

\section{\label{sec:introduction}Introduction}

Topological lattice paradigms have emerged as a physically intriguing framework for the 
manifestation of unprecedented wave phenomena across a broad range of settings, including condensed matter systems, photonics, acoustics, and mechanical metamaterials~\cite{Hasan2010,Qi2011}. A defining feature of such paradigms is that lattice geometry and symmetry can endow linear wave spectra with topologically protected characteristics, such as edge and interface modes that persist under perturbations and disorder. These developments have motivated extensive efforts to classify and realize topological phases in discrete systems, with particular emphasis on tight-binding and lattice-based descriptions \cite{Lu2014,Ozawa2019}.

On the other hand, 
 dispersive nonlinear lattice 
models have been a subject of extensive investigation for multiple decades~\cite{Aubry2006,Flach2008}. 
Relevant developments have been motivated, among numerous other areas,
from the study of 
optical waveguides~\cite{LEDERER20081}
(as well as of continuum periodic photorefractive media) and the exploration of mean-field
atomic Bose-Einstein condensates (BECs) in optical lattice
potentials~\cite{RevModPhys.78.179}. The latter setting is, in fact, enjoying an ever
increasing level of accessibility through very recent state-of-the-art experiments~\cite{elmar}.
Arguably, the discrete nonlinear 
Schr{\"o}dinger (DNLS) equation~\cite{kev09,chriseil} constitutes 
the most universal  model of this type of systems.
The study
of solitary waves, instability phenomena, and nonlinear dynamics in discrete
nonlinear optics~\cite{cole} as well as in atomic BECs~\cite{BrazhnyiKonotop2004MPLB} has, typically,
revolved around the mathematical, computational and experimental developments in 
connection to the DNLS model.

While many of the above topological nonlinear investigations have focused on 
one- and two-dimensional systems, the landscape of three-dimensional 
settings
at the nexus of dispersive nonlinearity and topological (flat band) lattices is
far more sparse; see, e.g.,~\cite{LiEtAl2022CommsPhys,LiLiJiaLiu2022PRB} for some
topological nonlinear waveform examples, including in systems such as three-dimensional (3d) photonic
Chern insulators \cite{Kattan2026}. 
An intriguing example of a topological lattice in three spatial dimensions consists
of the pyrochlore lattice.
This lattice belongs to the crystal space group Fd3m. It typically contains magnetic atoms
arranged to form a lattice of tetrahedral motifs joined at each corner. It is a prominent 
3d frustrated lattice structure and has been extensively studied in the exploration of
spin ice \cite{Ortiz24} as well as in the context of topological phases \cite{Ueda17}. Its 2d analog is the 
Kagom{e} lattice~\cite{Sutherland1986Localization}, which 
has been a subject of considerable investigation 
in its own right, e.g., in photonic contexts~\cite{avadhkagome},
due to intriguing features such as (linear and) nonlinear
compactly supported states and associated symmetry
breaking bifurcations~\cite{Vicencio13,shi2025stabilitytheoryflatband}. 

The study of Heisenberg-like antiferromagnets and Hubbard model on the pyrochlore lattice reveals such exotic
phases as quantum spin liquids \cite{Normand14,Normand16}, Dirac \cite{Wan11} and Weyl \cite{Sushkov15} semimetals as well as topological insulators \cite{Otsuka21} driven by competing interactions. The latter include spin-orbit coupling and electron correlations which in conjunction with the pyrochlore lattice’s inherent frustration lead to emergent phenomena, e.g., spin-charge separation and exotic excitations \cite{Savary16}. In materials science its tunability is particularly desirable because the relevant structure allows for incorporating a variety of materials that enable tuning of properties including in the now ubiquitous metal organics frameworks (MOFs) \cite{Nutakki23}.
From the solid state chemistry perspective, pyrochlores are mixed-metal oxide materials that have the
general formula $A_2B_2O_7$. Some important examples include $Gd_2Ti_2O_7$ , $Y_2Zr_2O_7$, 
$Bi_2Ru_2O_7$ and $Cd_2Re_2O_7$ \cite{Connor21} with varied applications in spin ice physics, quantum magnetism, superconductivity, colossal magnetoresistance, ceramics, thermal barrier coatings, fuel cells, etc.~\cite{Anantharaman21}.

Our focus in the present work will be at the intersection of the pyrochlore lattice and 
dispersive nonlinear settings as they are universally represented by the DNLS
model~\cite{chriseil,LEDERER20081,kev09,cole}. 
While the 2d Kagom{e} analogue
has been extensively studied~\cite{avadhkagome,Vicencio13,shi2025stabilitytheoryflatband}
(see also~\cite{kagome2,kagome3,kagome4}), and so have other 
related lattices such as the Lieb lattice~\cite{Vicencio2015LocalizedStates}
[see also reviews, such as~\cite{leykam2013flat,leykam2018artificial, danieli2024flat, rhim2021singular}].
{ A recent complementary asymptotic study of a nonlinear Kagom{e} lattice has also derived nonlinear-Schr{\"o}dinger-type reductions for weakly nonlinear localized modes, including an expansion near a flat-band/dispersive-band touching point~\cite{Wattis2026}.}
The 3d settings have been
far more sparse by comparison; an example of a standard cubic lattice example
appears, e.g., in~\cite{LUKAS2008339}. 
Both the Kagom{e} and Lieb lattice 
harbor flat bands~\cite{Kumar2025}.  In the magnetic context the Kagom{e} lattice is maximally frustrated whereas the Lieb lattice is unfrustrated \cite{LopezBezanilla2025}. Here, we aim to extend the interplay of
onsite nonlinearity and flat-band topology to the prototypical 3d 
DNLS pyrochlore setting.
While it is less obvious how to implement this at the nonlinear optical setting,
recent advancements in the realization of optical lattices in BEC~\cite{elmar}
and the associated remarkable control over relevant optical potentials through the so-called digital micromirror devices~\cite{Gauthier2016DMDOpticalPotentials} suggest 
that this setting may very soon be experimentally accessible, in addition to being
of theoretical and computational interest. 

In what follows, in section II we present
the relevant DNLS model and the corresponding theoretical setup, including
the analysis of the linear system, its dispersion characteristics and how
these shape the potentially accessible nonlinear states.
We provide theoretical considerations of such states including the most
fundamental single site solitons, the Kagom{e} (as well
as the honeycomb/triangular~\cite{LawKevrekidisKoukouloyannisKourakisFrantzeskakisBishop2008PRE})
lattice-inspired vortices of topological charge $S=1$ and $S=2$, and
equally centrally to this flat band lattice, the compactly supported
nonlinear states. The latter stem from the linear limit but persist in the nonlinear
realm and understanding their existence and stability properties in the
presence of nonlinearity is of intrinsic interest~\cite{sandra}. In section III
we provide numerical results for each of these states, both in the focusing,
and in the defocusing nonlinear realm. Here, we bear once again in mind the
setup of~\cite{elmar} where quenches from one regime to the other ---e.g., from 
focusing to defocusing or vice versa--- are routinely accessible.
Finally, section IV summarizes our main findings and provides a number of avenues for future
work.

\section{\label{sec:model}Model and Theoretical Setup}
We will consider the three-dimensional nonlinear lattice 
dynamical equation:
\begin{align}
    {i\dot{u}_{n}+C(Au)_{n}+\gamma|u_{n}|^2u_{n}} & {=0.}
\end{align}
{Here $C$ is the coupling strength, $A$ is the (negative) nearest-neighbor graph Laplacian of the pyrochlore lattice{, and $\gamma$ is the strength of the nonlinearity}.  Since every bulk pyrochlore site has six nearest neighbors, $A$ is specified explicitly by
\begin{align}
 A_{nm}=\begin{cases}
 -6, & n=m,\\
 1, & m\in\mathrm{NN}(n),\\
 0, & \text{otherwise},
 \end{cases}
 \qquad
 (Au)_n=\sum_{m\in\mathrm{NN}(n)}(u_m-u_n),
 \label{eq:pyro-laplacian}
\end{align}
so that each row sums to zero; equivalently, $A=G-6I$, where $G$ is the nearest-neighbor adjacency matrix.  This is the standard nearest-neighbor pyrochlore connectivity; see, e.g., Ref.~\cite{Bergman08}. Here, the set $\mathrm{NN}(n)$ represents the nearest neighbors of
the given site $n$ within the pyrochlore lattice}

{We also clarify the indexing and crystallographic convention used throughout the manuscript in what follows. The subscript $n$ is a single global site label, rather than a Cartesian coordinate. More explicitly, a pyrochlore site may be indexed as $n\equiv(\mathbf R,\alpha)$, where $\mathbf R=j_1\mathbf a_1+j_2\mathbf a_2+j_3\mathbf a_3$ labels an FCC Bravais cell and $\alpha\in\{0,1,2,3\}$ labels the four sites of the pyrochlore basis. Taking $a_0$ to be the conventional cubic lattice constant, we use the primitive vectors
\begin{align}
 \mathbf a_1&=\frac{a_0}{2}(0,1,1), &
 \mathbf a_2&=\frac{a_0}{2}(1,0,1), &
 \mathbf a_3&=\frac{a_0}{2}(1,1,0),
 \label{eq:pyro-primitive-vectors}
\end{align}
and the four-site basis
\begin{align}
 \boldsymbol\rho_0&=(0,0,0), &
 \boldsymbol\rho_1&=\frac{\mathbf a_1}{2}, &
 \boldsymbol\rho_2&=\frac{\mathbf a_2}{2}, &
 \boldsymbol\rho_3&=\frac{\mathbf a_3}{2}.
 \label{eq:pyro-basis}
\end{align}
The four basis sites form one tetrahedral primitive cell. In the numerical implementation the pair $(\mathbf R,\alpha)$ is flattened to $n=1,\ldots,N$. Correspondingly, the Bloch ansatz can be written as $v_{\mathbf R,\alpha}=V_\alpha(\mathbf k)e^{i\mathbf k\cdot\mathbf R}$. The wave vector $\mathbf k$ has three independent components; in the dispersion relation below we use the Cartesian components $(k_x,k_y,k_z)$ (equivalently, one may use coordinates in the reciprocal primitive basis). Thus the three-dimensional character is fully retained even though a single enumerating site label $n$ is used in the real-space DNLS equation.}

{ It is relevant to add here that the origin of the flat band also admits a simple real-space interpretation. The pyrochlore lattice is the line graph of the diamond lattice. On an elementary hexagonal loop, one may assign alternating amplitudes so that at every neighboring site outside the loop the contributions from the two adjacent loop sites cancel {\it exactly}. This destructive interference prevents leakage from the loop and produces a compact localized eigenstate. Translations and linear combinations of such loop states span the macroscopically degenerate flat-band subspace; for the nearest-neighbor pyrochlore model considered herein this yields the two degenerate flat bands discussed below~\cite{Bergman08}.}

\begin{figure}[!tbp]
    \centering
    \includegraphics[width=0.85\linewidth]{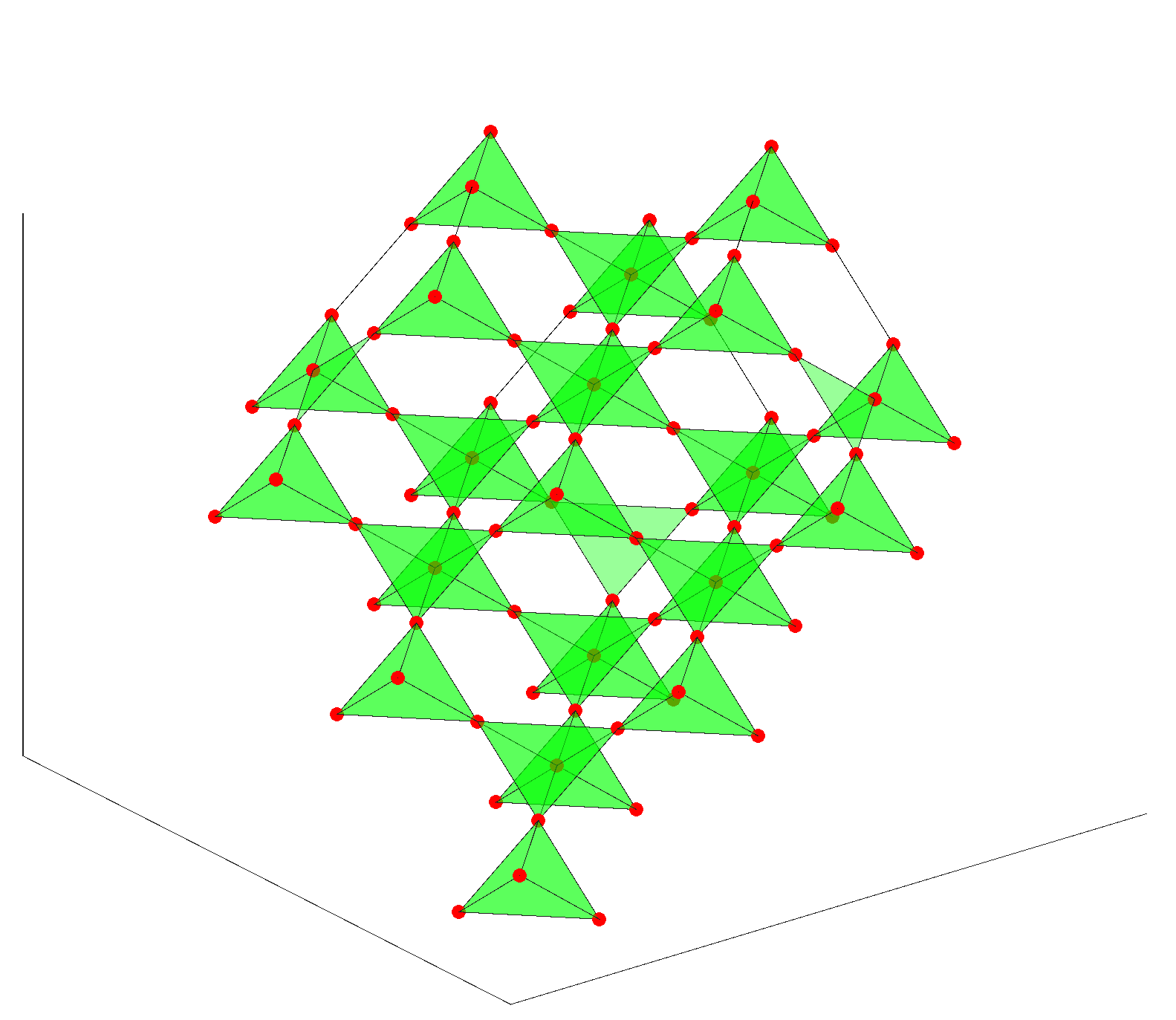}
    \caption{{Pyrochlore lattice with nearest-neighbor links. The lattice is an FCC Bravais lattice with a four-site tetrahedral basis; the primitive vectors and basis positions used here are given in Eqs.~\eqref{eq:pyro-primitive-vectors}--\eqref{eq:pyro-basis}. The 3d structure is formed by corner-sharing tetrahedra (green).}}
    \label{fig:pyrochlore-lattice}
\end{figure}

We can observe the basic spatial structure of this lattice, with the associated couplings in Fig. \ref{fig:pyrochlore-lattice}.

\subsection{Linear spectrum}
Before turning to the nonlinear states that the lattice can
support, for completeness, we briefly remind the reader
of the lattice's linear band structure.
In particular, we  consider the standard standing wave 
Ansatz $u_{n}(t)=v_{n} e^{i\mu t}$. Then, $v_{n}$ satisfies the following equation:
\begin{align}\label{eq:dnls}%
    {\mu v_{n}=C(Av)_n+\gamma|v_{n}|^2v_{n}.}
\end{align}

In the absence of the nonlinearity (e.g., for $\gamma \rightarrow 0$
or for $|v_n| \rightarrow 0$), 
it is known that this lattice exhibits two degenerate flat bands at the largest (in absolute value) energy
level, in our case $E=-8 C$, and that this leads to compactly
supported eigenstates of the Laplacian that extend exactly to nonlinear states \cite{Bergman08}.

\begin{figure}[!tbp]
    \centering
    \begin{subfigure}[b]{0.45\columnwidth}
        \centering
        \includegraphics[width=\linewidth]{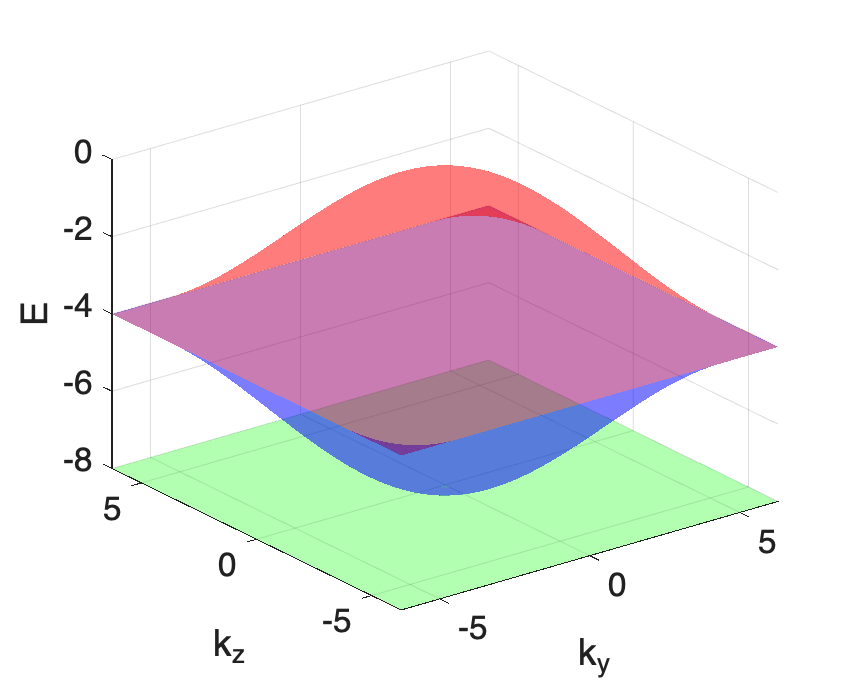}
    \end{subfigure}%
    ~
    \begin{subfigure}[b]{0.45\columnwidth}
        \centering
        \includegraphics[width=\linewidth]{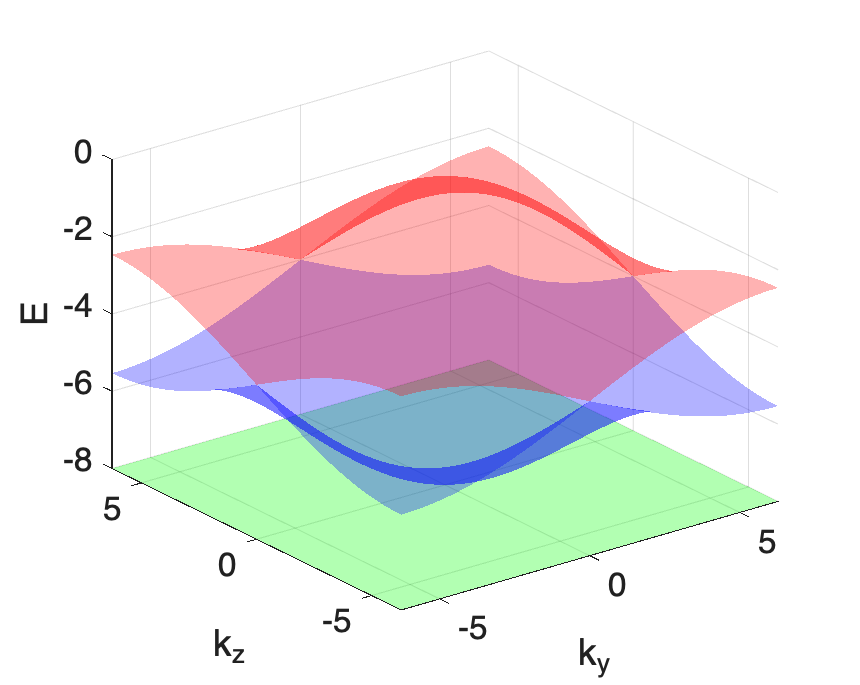}
    \end{subfigure}
    \begin{subfigure}[b]{0.45\columnwidth}
        \centering
        \includegraphics[width=\linewidth]{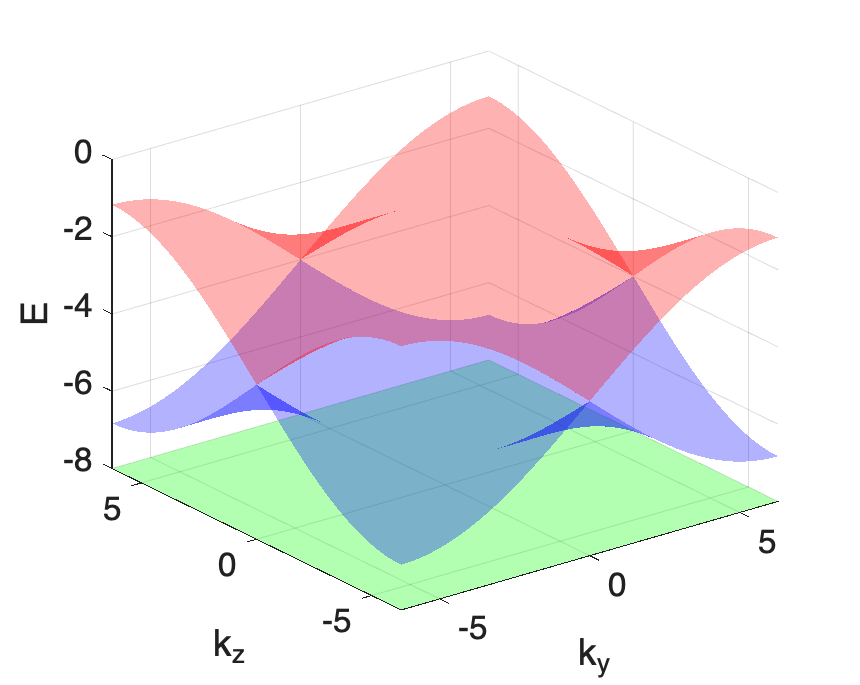}
    \end{subfigure}%
    ~
    \begin{subfigure}[b]{0.45\columnwidth}
        \centering
        \includegraphics[width=\linewidth]{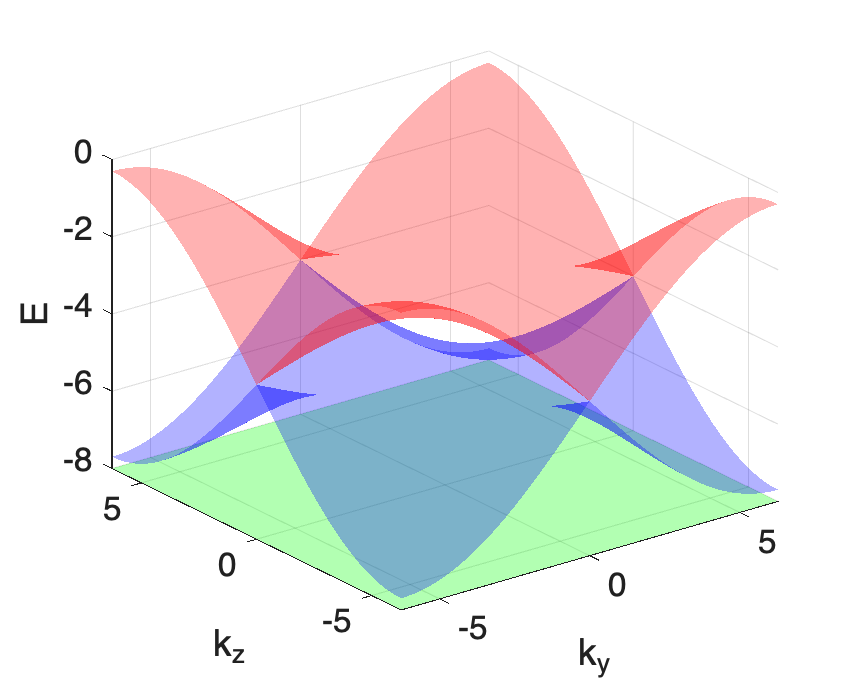}
    \end{subfigure}
    \caption{Dispersive and flat bands for $C=1$ and for different values of {$k_x$}: (Top left) {$k_x=0$}; (Top right) {$k_x=\pi/2$}; (Bottom left) {$k_x=\pi$}; (Bottom right) {$k_x=3\pi/2$}.}
    \label{fig:band_structure}
\end{figure}

Additionally, this lattice exhibits two dispersive bands at the energy levels:
\begin{widetext}
    \begin{align}
        {E_{\pm}(k_x,k_y,k_z)} & {=2C \left[-2\pm \sqrt{1+\cos\left(\frac{k_x}{2}\right)\cos\left(\frac{k_y}{2}\right)+\cos\left(\frac{k_y}{2}\right)\cos\left(\frac{k_z}{2}\right)+\cos\left(\frac{k_z}{2}\right)\cos\left(\frac{k_x}{2}\right)}\right],}
    \end{align}
\end{widetext}
{where $(k_x,k_y,k_z)$ are the Cartesian components of the three-dimensional Bloch wave vector $\mathbf k$ (in units consistent with the lattice convention above).} We can
observe this band structure in Fig. \ref{fig:band_structure}, where since $E$ is
{symmetric in $k_x,k_y,k_z$ we may for clarity fix $k_x$ and visualize $E$ as a function of $k_y,k_z$.}

Since the flat band eigenstates can be calculated exactly \cite{Bergman08}, they are of particular interest and will
be a core part of our study. In addition to these states, we will also look at nonlinear waveforms such as single-site and compactly 
supported
solitary waves, as well as vortices that appear in this system. In the first two cases, we will take $v_n$
to be real, while in the case of vortices we will consider $v_n$ to be complex. Following the path discussed, e.g., in~\cite{kev09},
we will first seek to explore conditions for the persistence
of such states.

Once the persistence of these nonlinear wave patterns is verified,
we will also look at the spectral stability analysis of the resulting states.
{For clarity, we spell out the linearization leading to the operators $L_+$ and $L_-$.  For a real stationary state $v_n$, set
\begin{align}
 u_n(t)=e^{i\mu t}\big[v_n+a_n(t)+i b_n(t)\big],
\end{align}
and retain only terms that are linear in the real perturbations $a$ and $b$.  Using the stationary equation gives
\begin{align}
 \dot a=L_-b,\qquad \dot b=-L_+a,
 \label{eq:linearized-real-imag}
\end{align}
where
\begin{align}\nonumber
 L_+&=-CA+\mu I-3\gamma\,\mathrm{diag}(v_n^2),\\
 L_-&=-CA+\mu I-\gamma\,\mathrm{diag}(v_n^2).
\end{align}
Eliminating $a$ from Eq.~\eqref{eq:linearized-real-imag} and taking $b(t)=e^{\lambda t}w$ yields
\begin{align}
 \lambda^2 w=-L_+L_-w,
\end{align}
which is the eigenvalue problem used below. {This real--imaginary decomposition and the associated $L_+$ and $L_-$ operators are key as
concerns the spectral linearization of NLS and DNLS standing waves; see, e.g., Refs.~\cite{SulemSulem1999,kev09,Carretero24}.} Spectral stability requires the resulting temporal eigenvalues $\lambda$ to have a vanishing real part.}

\subsection{Rescaled equation: Existence and Stability}

We start by rescaling the equation using $v_n=|\mu|^{1/2}w_n$, and $\tilde{C}=C/|\mu|$. This gives us the following equation:
\begin{align}
    \tilde{C}Aw_n+(\gamma|w_n|^2-\text{sgn}(\mu))w_n & =0,
\end{align}
where $\text{sgn}(\mu)$ is the sign of $\mu$. For simplicity, we will 
use the symbols $\varepsilon$ instead of $\tilde{C}$, and $v$ instead of $w_n$,
so that the rescaled equation is:
\begin{align}\label{eq:rescaled_dnls}%
    \varepsilon Av+(\gamma|v|^2-\text{sgn}(\mu))v & =0.
\end{align}

We can look at the different states again in this new formulation, starting from the anti-continuum (AC)
limit, $|\mu|\gg 1$ and $\varepsilon\sim|\mu|^{-1}\approx 0$. 
This is a limit that in earlier works has enabled the
derivation of solvability conditions associated with the
persistence and the stability of different multi-site 
nonlinear states~\cite{kev09}.
This is because at the AC limit, sites can either 
possess a vanishing amplitude or be of the form
$v_n=\sqrt{\text{sgn}(\mu)/\gamma} e^{i \theta_n}$,
where $\theta_n$ is an arbitrary phase.

For the compactly supported states, we may consider the $L_{\pm}$ operators:
\begin{align}\nonumber 
    L_+ & =-\varepsilon A+(\text{sgn }(\mu)-3 \gamma v^2), \\
    L_- & =-\varepsilon A+(\text{sgn }(\mu)-\gamma v^2).
\end{align}
Following the same procedure as in \cite{kev09},
in the AC limit we may reduce the eigenvalue problem for the full equation (i.e. $\lambda^2w= -L_+L_-w$) to the following reduced problem:
\begin{align}\label{eq:reduced_eigenvalue_problem}
    \lambda^2 w = 2\left\langle w, L_-w \right\rangle.
\end{align}
Thus, we essentially need to compute the eigenvalues of $L_-$.
Then, Eq.~(\ref{eq:reduced_eigenvalue_problem}) provides us 
with a prescription as to how to ``augment'' these eigenvalues
into eigenvalues of the full nonlinear problem.

More generally, for both the compactly supported states and the vortices,
we may {make use of the Lyapunov-Schmidt solvability condition:}
\begin{align}
    g_n&=\sum_{m\in \mathrm{NN}(n)}\sin(\theta_n-\theta_m)=0,
\end{align}
{where the summation subscript denotes that $m$ and $n$
are nearest neighbors (i.e., $A_{mn}=1$). Expanding in powers of $\varepsilon$ in \eqref{eq:reduced_eigenvalue_problem} we may obtain a closed form expression for the eigenvalues of the linearization near the AC limit. For the focusing case, this yields the eigenvalue problem:}

\begin{align}
    \lambda^2     & = 2 \varepsilon \alpha, \\
    \mathcal{M} c & = \gamma\alpha c,
\end{align}
where $\mathcal{M}$ {is the Jacobian of the solvability conditions, i.e., $\mathcal{M}_{ij}=\partial g_i/\partial\theta_j$. Since we consider states supported only on six sites, and with $\theta_m-\theta_n\equiv \Delta \theta$ for all $n, m$, this} is a $6\times 6$ tridiagonal matrix, with 
the vector of sub-diagonal, diagonal and super-diagonal terms reading:
$(-\cos\Delta\theta,2\cos\Delta\theta,-\cos\Delta\theta)$.
The resulting matrix
has eigenvalues $\gamma\alpha=0, \cos\Delta\theta, 3\cos\Delta\theta, 4\cos\Delta\theta$, with $\cos\Delta\theta$ and
$3\cos\Delta\theta$ being doubly degenerate.
Thus, the eigenvalues of the linearization near the AC limit are given by:
\begin{align}\label{lam1}
    \lambda & = 0,                                         \\
    \lambda & = \pm \sqrt{\gamma\varepsilon \cos\Delta\theta},  \\
    \lambda & = \pm \sqrt{3\gamma\varepsilon \cos\Delta\theta}, \\
    \lambda & = \pm \sqrt{4\gamma\varepsilon \cos\Delta\theta}.\label{lam2}
\end{align}

Therefore {in the focusing case ($\gamma=1$)}, if $\cos\Delta\theta>0$ ---as is the case, e.g.,
for $\Delta \theta=\pi/3$, these eigenvalues will be real, and the state will be unstable. Otherwise,
they will be pure imaginary, and the state will be stable near the AC limit. {Conversely, in the defocusing case ($\gamma=-1$), if $\cos\Delta\theta>0$, the state would be stable near the AC limit, and if $\cos\Delta\theta<0$ then the state will be unstable.}

\subsection{Single site soliton}
For the single-site soliton solution in the focusing ($\gamma=1$) case, we recall the Vakhitov-Kolokolov criterion for stability
\cite{Vakhitov73}, which establishes that for $P$ the power of the state, the condition
for stability is:
\begin{align}\label{eq:vakhitov}
    \frac{dP}{d\mu} & > 0.
\end{align}
Thus, we expect the onset of instability to occur at a 
(non-degenerate) local extremum of $P$, where the relevant
monotonicity changes.

\subsection{Compactly Supported State}
Following \cite{Bergman08, Vicencio13}, we find an exact solution of the
linear system in the form of a ring, with six peaks of equal amplitude and alternating sign. We can observe this configuration in Fig. \ref{fig:cls} below.

\begin{figure}[!tbp]
    \centering
    \includegraphics[width=.95\linewidth]{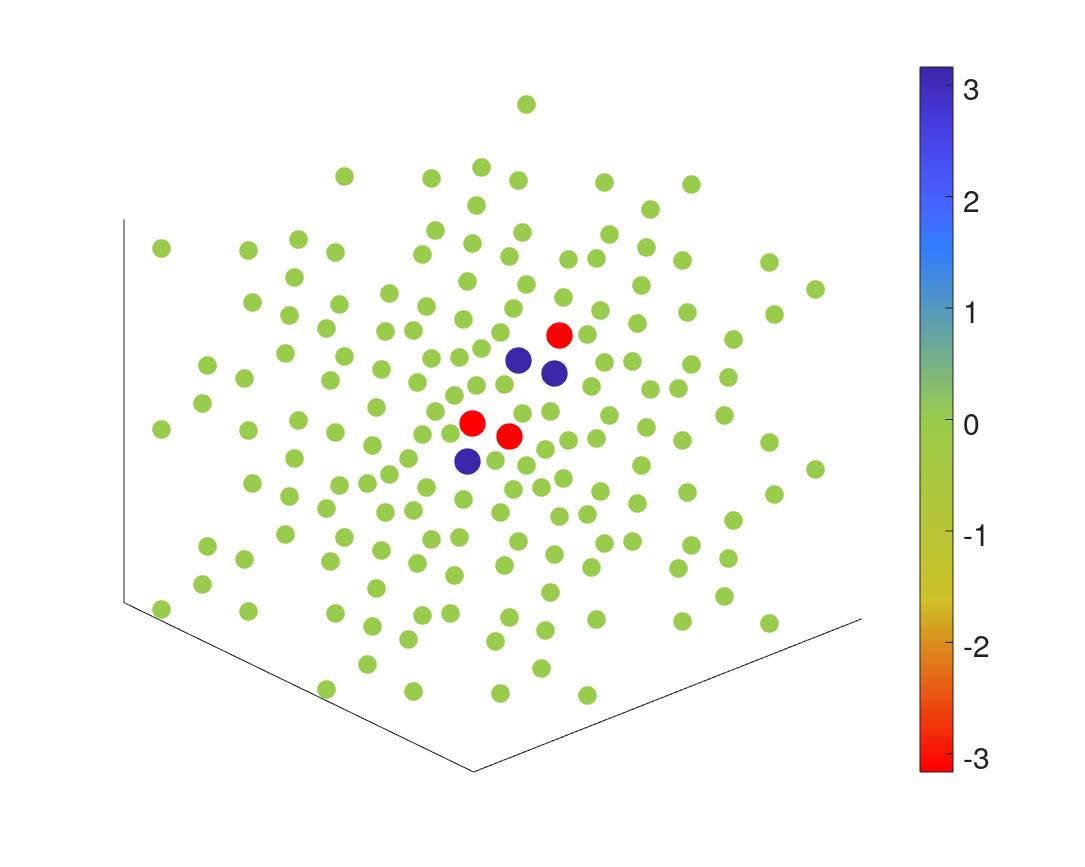}
    \caption{A compactly supported state for $\mu=2$. Excited sites look bigger.}
    \label{fig:cls}
\end{figure}

This can be continued to a nonlinear state in the exact same configuration, with power $P=\sum_n |v_n|^2$. Since this state is supported only on six sites, we can solve for this exactly in \eqref{eq:dnls} as a function of $\mu$, obtaining:
\begin{align}
    P & =\frac{6(8+\mu)}{\gamma}.
\end{align}
We note that, since $P\geq 0$, in the focusing ($\gamma=1$) case we must have $\mu\geq -8$, and in the defocusing ($\gamma=-1$) case we must have $\mu\leq -8$. For simplicity, $C=1$ here.

Next, we consider the rescaled equation labeled as \eqref{eq:rescaled_dnls}.
We can consider the anti-continumm (AC) limit, corresponding to $\varepsilon\approx0$, or equivalently
to $|\mu|\gg 1$, and look at the stability of the compactly supported states near this limit. We
first consider the eigenvalue problem \eqref{eq:reduced_eigenvalue_problem}.
We consider both focusing ($\gamma=1$, $\mu\gg 1\Rightarrow\text{sgn }(\mu)=1$),
and defocusing ($\gamma=-1$, $-\mu\gg 1\Rightarrow\text{sgn }(\mu)=-1$) cases. Additionally, since we are considering compactly supported states, we have an explicit expression for $P$:
\begin{align}
    P & =\frac{6\left(\text{sgn}(\mu)+8\varepsilon\right)}{\gamma},
\end{align}
and consequently $v^2=(\text{sgn}(\mu)+8\varepsilon)/\gamma$ on the six excited sites.

{An important distinction from the setting considered in Ref.~\cite{shi2025stabilitytheoryflatband} is that the pyrochlore lattice has two flat Bloch bands at the same energy, whereas the stability theory developed there assumes a single flat band at the relevant spectral edge.  The resulting flat-band eigenspace is therefore larger in the present problem.  The particular six-site alternating-sign compact state studied below selects one specific element of this eigenspace and continues exactly into the nonlinear problem because its 
alternating sites preserve the destructive-interference cancellation.  Thus, the additional flat-band degeneracy does not alter the explicit existence relation for this chosen compact branch, but it does enlarge the space of flat-band perturbations and linear combinations and must be accounted for in a general small-amplitude stability theory.  For this reason, the criterion of Ref.~\cite{shi2025stabilitytheoryflatband} cannot be applied verbatim to the pyrochlore case, but rather constitutes an interesting extension
for potential future study.}

In addition to this, we may also compute the eigenvalues of $\mathcal{M}$ from Eqs.~(\ref{lam1})-(\ref{lam2}) to estimate how
the (small imaginary or real) eigenvalues of the linearization behave near the AC limit. In both
cases, this will give us the $\mathcal{O}(\sqrt{\varepsilon})$ behavior of the eigenvalues.
Since we have $\Delta \theta=\pi$ for the compactly supported states, and $\cos\pi=-1$,
we find that the eigenvalues of the linearization {in the focusing case} are imaginary, and they will grow near the AC limit as:
\begin{align}\nonumber
    \lambda\sim \pm i\sqrt{2\varepsilon},\\\nonumber
    \lambda\sim \pm i\sqrt{6\varepsilon},\\\label{eq:evalues-cls-theoretical}
    \lambda\sim \pm i\sqrt{8\varepsilon}.
\end{align}
Recall that we expect the onset of instability to occur when one of these eigenvalues collides with the continuous spectrum, starting at $\lambda=i$. This gives us an estimate for the onset of instability at $\varepsilon=\frac{1}{8}$, or equivalently $\mu=8$.

{Conversely, in the defocusing case these eigenvalues are positive real, and they grow as:}
\begin{align}\nonumber
    \lambda\sim \pm \sqrt{2\varepsilon},\\\nonumber
    \lambda\sim \pm \sqrt{6\varepsilon},\\\label{eq:evalues-cls-theoretical-defocusing}
    \lambda\sim \pm \sqrt{8\varepsilon}.
\end{align}
{This suggests that the compactly supported state is unstable for any value of $\varepsilon$, or equivalently for any $\mu$ in the vicinity
of the AC limit.}

\subsection{Vortices}
Proceeding as for the compactly supported states in the vicinity of 
the AC limit, we may compute the eigenvalues of $\mathcal{M}$ to estimate how the eigenvalues of the linearization behave near the AC limit, as per Eqs.~(\ref{lam1})-(\ref{lam2}). We consider both charge-1 and charge-2 vortices, first off in the focusing case:

\begin{itemize}
    \item Charge-1 vortex: here we have $\Delta \theta=\pi/3$
    for a $2\pi$ phase winding over the 6-site contour, and 
    therefore $\cos(\pi/3)=1/2$. Accordingly, the eigenvalues of the linearization
          are real, and they will grow as:
          \begin{align}\nonumber
              \lambda\sim \pm \sqrt{\varepsilon},\\\nonumber
              \lambda\sim \pm \sqrt{3\varepsilon},\\\label{eq:evalues-vortex1-theoretical}
              \lambda\sim \pm \sqrt{4\varepsilon}.
          \end{align}
          In particular, this implies that the unit charge state is unstable for any $\varepsilon>0$ (i.e., for any $\mu>0$).
          This is in line with the earlier considerations
          in Kagom{e} lattice~\cite{avadhkagome}.

    \item Charge-2 vortex: for this case, since we have $\Delta \theta=2\pi/3$, and $\cos(2\pi/3)=-1/2$, the eigenvalues of the linearization
          are imaginary, 
          and their imaginary part will grow as 
          \begin{align}\nonumber
              \lambda\sim \pm i\sqrt{\varepsilon},\\\nonumber
              \lambda\sim \pm i\sqrt{3\varepsilon},\\\label{eq:evalues-vortex2-theoretical}
              \lambda\sim \pm i\sqrt{4\varepsilon}.
          \end{align}
          Recall again that we expect the onset of instability to occur when the largest of these eigenvalues collides with the continuous spectrum, starting at $\lambda=i$. This gives us an estimate for the onset of instability at $\varepsilon=\frac{1}{4}$, or equivalently $\mu=4$ for
          the focusing charge-2 vortices.

\end{itemize}

It is important to note that in the defocusing case, the
eigenvalue picture gets fully reversed, namely the real
eigenvalues of unit charge become imaginary ones, while
the imaginary eigenvalues of charge-2 become real. Consequently,
in the defocusing realm it is the unit charge vortex that
is spectrally stable, while the doubly-charged one that is 
spectrally (and dynamically) unstable. 

\section{Numerical Results}

\subsection{Single site soliton}
First, we consider a single-site soliton. We start by considering the anti-continuum limit ($C=0$ or equivalently $\varepsilon=0$), and $v_n=0$ for all but one of the lattice nodes $n$. For $v_n\neq 0$, we can solve for it in terms of $\mu$ (in the context of Eq.~(\ref{eq:dnls})), obtaining:
\begin{align}
    |v_n|^2 & =\frac{\mu}{\gamma}.
\end{align}
We can observe that, in a way analogous to the previous case, in the focusing ($\gamma=1$) case we must have $\mu\geq 0$, and in the defocusing ($\gamma=-1$) case we must have $\mu\leq 0$. In both cases, we will continue until $C=1$ for one value of $\mu$, and then consider the $P$ vs $\mu$ curve by continuing in $\mu$.

For the focusing case, we continue from $\mu=4$ to $\mu=0$, and look at the $P$ vs $\mu$ curve.
For the stability of the states as a function of $\mu$, we can numerically track
the magnitude of the eigenvalues of the linearization with the largest real part.
We can plot these two curves ($P=P(\mu)$ and { the} largest real
$\lambda$ as a function of $\mu$) together to check the theoretical predictions from the Vakhitov-Kolokolov criterion, as the curve
for $P$ changes monotonicity thereafter (around $\mu\gtrsim 1.12$), with the decreasing
portion pertaining to a spectrally unstable solitary wave.
This also connects (through the presence of a minimal $P$)
with the well-established theory of excitation thresholds
for the solitary waves in higher-dimensional DNLS 
lattices~\cite{MIWeinstein_1999}.

\begin{figure}[!tbp]
    \centering
 \begin{subfigure}[b]{0.95\columnwidth}
     \centering
  \includegraphics[width=\linewidth]{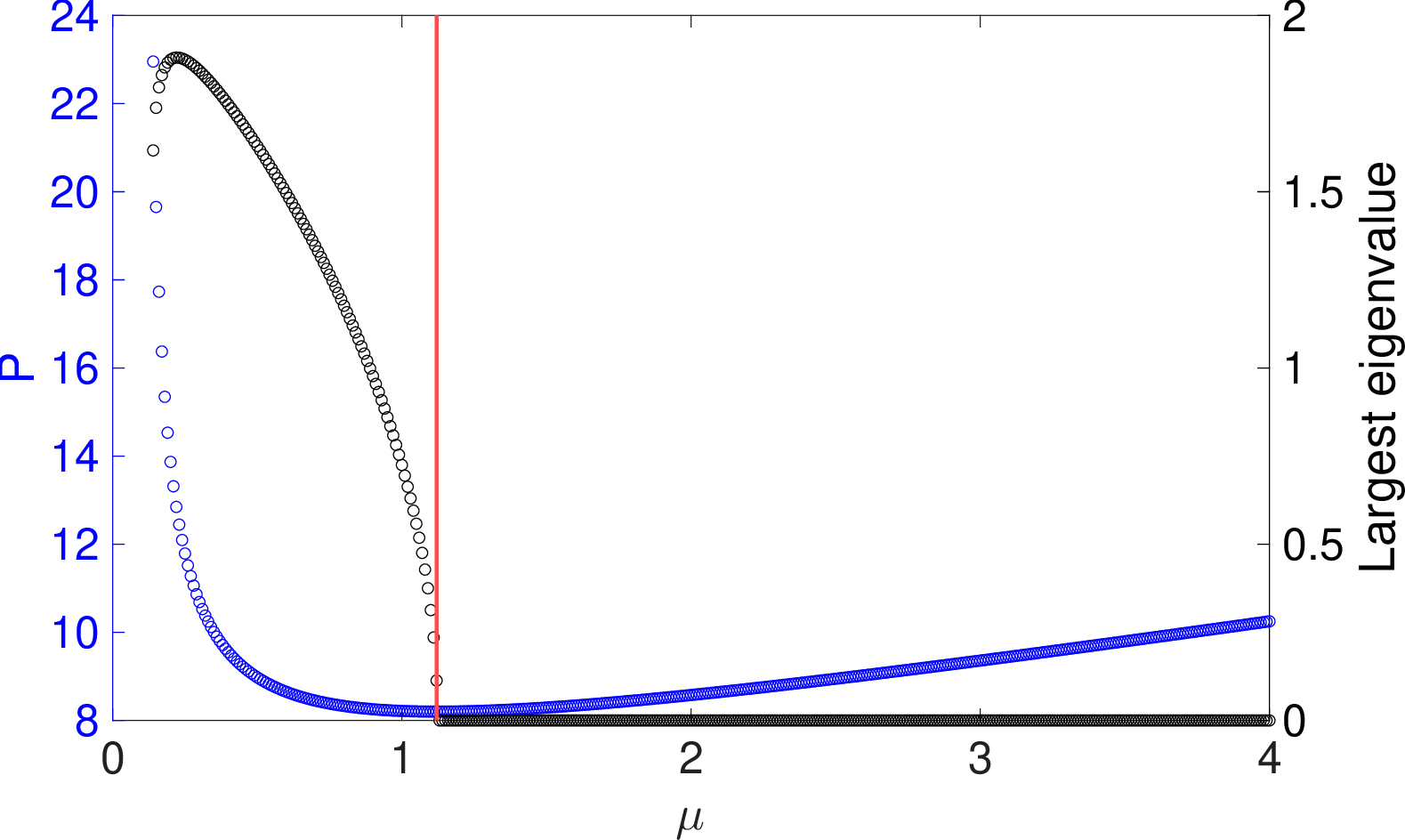}
  \end{subfigure}%
  \\
  \begin{subfigure}[b]{0.95\columnwidth}
    \centering
    \includegraphics[width=\linewidth]{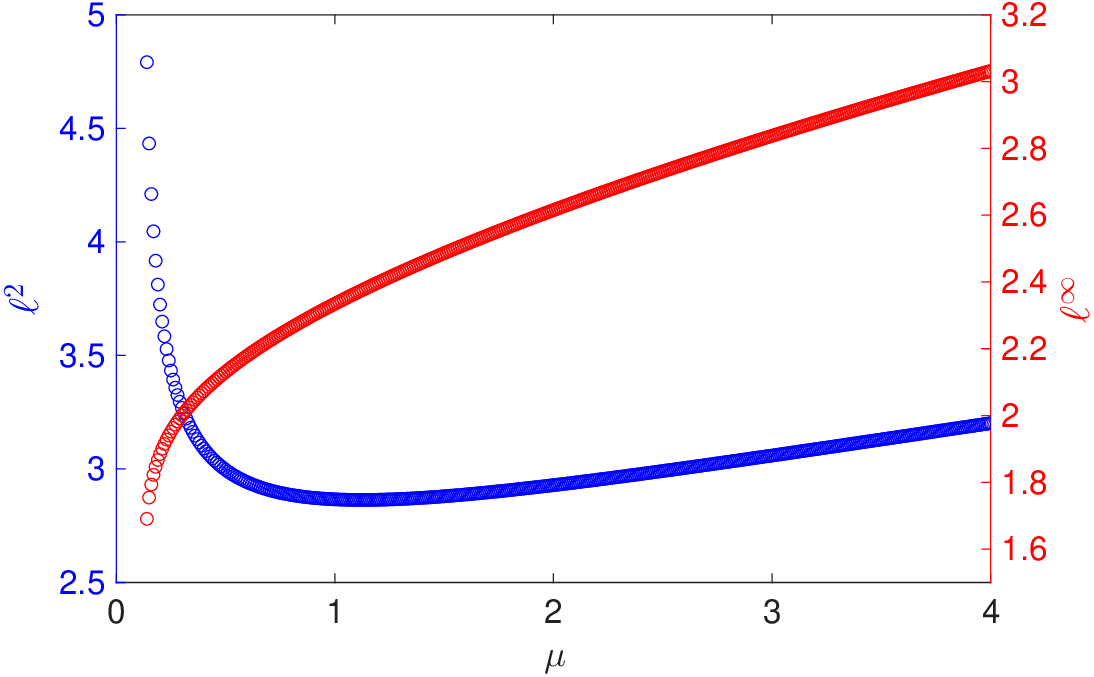}
  \end{subfigure}
    \caption{{Top figure: Power of the single-site state $P$ vs $\mu$ curve (left axis, in blue), and largest eigenvalue of the linearization (right axis, in black), focusing case. The red line marks the minimum of the $P$ vs $\mu$ curve. Bottom figure: $\ell^2$-norm (i.e., $\sqrt{P}$, left axis, in blue), and $\ell^\infty$-norm (right axis, in red) of the single-site state, focusing case.}}
    \label{fig:evalues-soliton-focusing}
\end{figure}

\begin{figure}[!tbp]
 \centering
 \begin{subfigure}[b]{0.9\columnwidth}
     \centering
  \includegraphics[width=\linewidth]{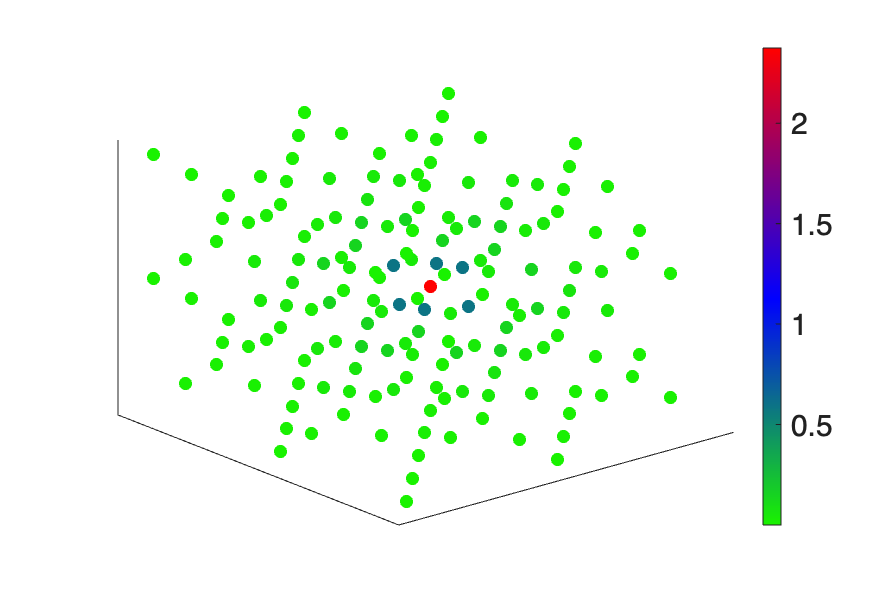}
  \end{subfigure}%
  \\
  \begin{subfigure}[b]{0.9\columnwidth}
    \centering
    \includegraphics[width=\linewidth]{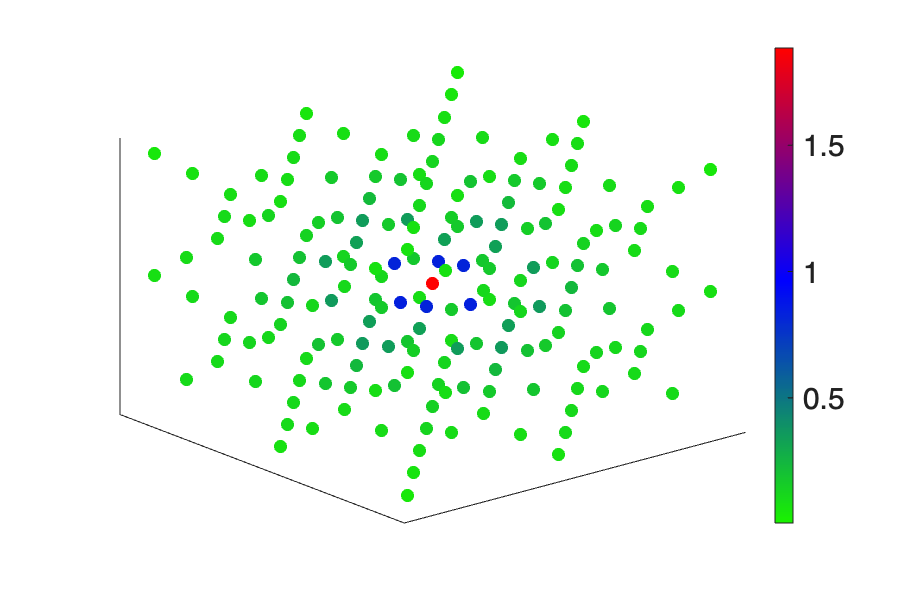}
  \end{subfigure}
 \caption{{ Profile of the focusing soliton at the minimum of the power $P(\mu)$ ($\mu\approx 1.12$, top), and below the minimum ($\mu=0.2$, bottom).}}
 \label{fig:soliton-min}
\end{figure}

{ It is worth noting that the soliton profile starts spreading out across the entire lattice as $\mu\downarrow 0$, as can be seen from the decaying $\ell^\infty$ norm of the state in Fig. \ref{fig:evalues-soliton-focusing}. This is consistent with the validity of a continuum NLS approximation in this -small $\mu$- regime, similar to what was shown in \cite{hofstrand25} for the Kagome lattice. Moreover, at the minimum of the power ($\mu\approx 1.12$) we can see (cf. Fig. \ref{fig:soliton-min}) the soliton is broadly localized around a single lattice site, and it starts spreading out across the lattice as it becomes unstable.}

Similarly, for the defocusing case, we continue from $\mu=-12$ to $\mu=-8$, and look at the $P$ vs $\mu$ curve. 
{ It is important to recall here that this single-site branch
bifurcating from the regular quadratic band at $\mu=-8$ is {\it not}
compactly supported (as the hexagonal state bifurcating from the
flat band below), but rather exponentially localized.}
\begin{figure}[!tbp]
    \centering
    \includegraphics[width=.95\linewidth]{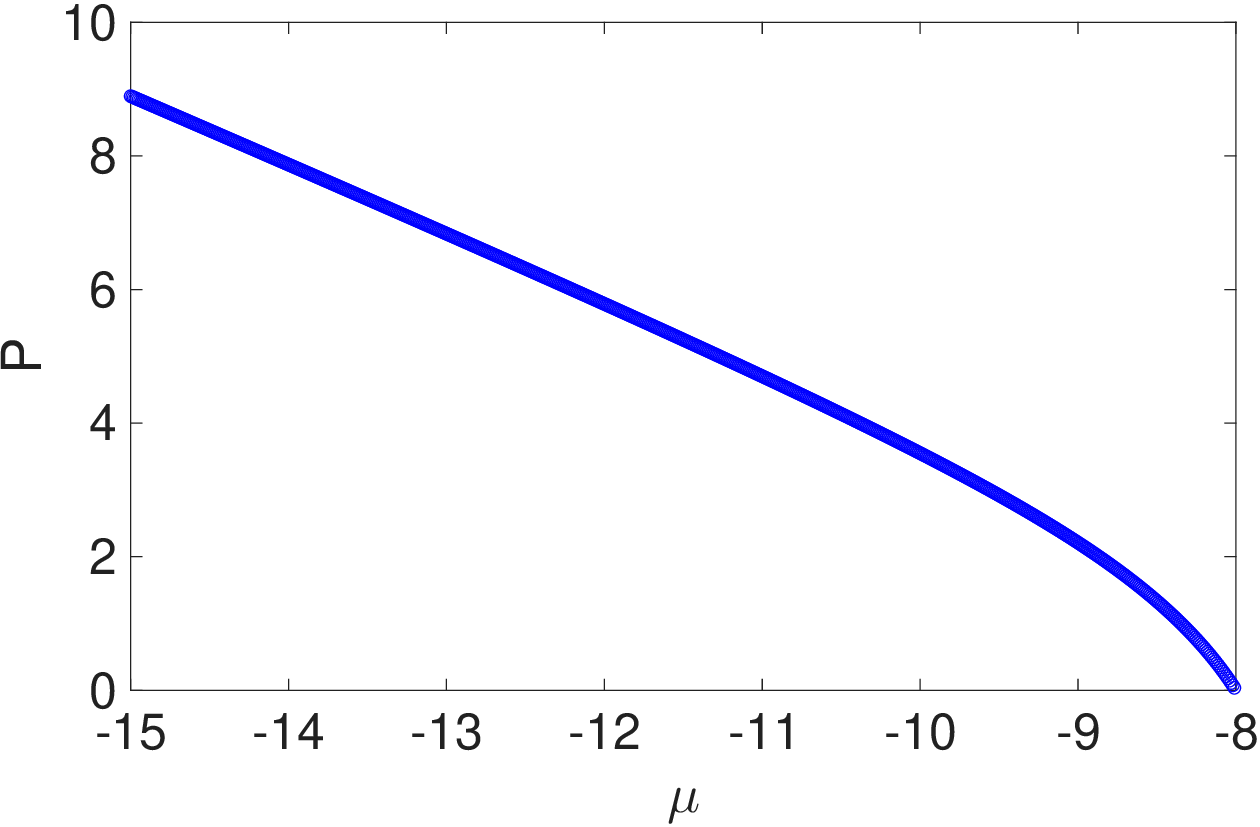}
    \caption{Power of the single-site state $P$ vs $\mu$ curve, defocusing case.}
    \label{fig:pvl-soliton-defocusing}
\end{figure}
We can observe the results in Fig. \ref{fig:pvl-soliton-defocusing}.
As before, for the stability of the states as a function of $\mu$, we can numerically track the magnitude of the eigenvalue of the linearization with the largest real part.
The results obtained suggest that the soliton is stable for $\mu<-8$,
{ The corresponding monotonicity for the defocusing case
suggests that the stability does not change due to a change in the sign of the slope as in the focusing case. Moreover, contrary, e.g., to the
Kagome case of~\cite{Vicencio13}, this branch does {\it not} collide
with any asymmetric branch toward a stability-changing bifurcation.
Rather, in a way partially reminiscent, e.g., of the checkerboard
lattice in Fig.~4 of the Supplement in~\cite{shi2025stabilitytheoryflatband},
the single site branch preserves its stability in this defocusing case for
all the values of $\mu$ considered.
}

\subsection{Compactly Supported States}
We will consider the results both from the original equation \eqref{eq:dnls}, and from the rescaled
equation \eqref{eq:rescaled_dnls}.

In so doing, we will examine the continuation of
\eqref{eq:dnls} over the variation of $\mu$ starting
from the linear limit. We will also examine the 
continuation of \eqref{eq:rescaled_dnls} from the opposite
end and the anticontinuum limit of $\varepsilon=0$.

\subsubsection{DNLS equation}
\begin{itemize}
    \item \textbf{Focusing case}

Recall that for this state the existence results are
``exact'' given the equal and opposite amplitudes of 
solely the 6 excited adjacent sites within the lattice
which change as a function of $\mu$ (or $\varepsilon$),
modifying accordingly the mode power.
          For the stability of the states as a function of $\mu$, we can numerically track the magnitude of the eigenvalues of the linearization with the largest real part.
          This can be seen in Fig. \ref{fig:evalues-CLS-focusing-rescaled-largemu}.

          \begin{figure}[!tbp]
              \centering
              \includegraphics[width=.95\linewidth]{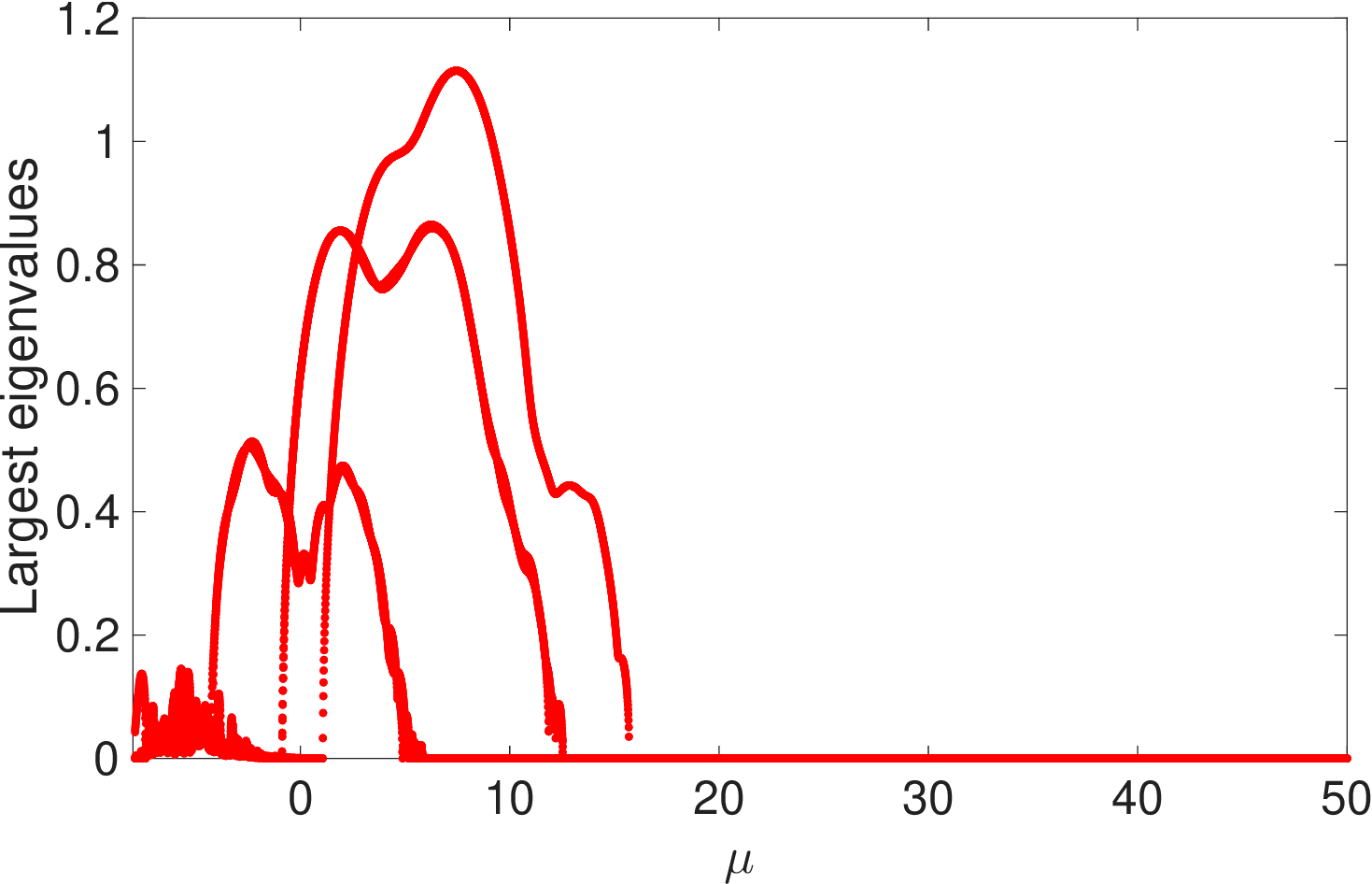}
              \caption{Positive eigenvalues of the linearization
              around the compactly supported state in the 
              focusing case.}
              \label{fig:evalues-CLS-focusing-rescaled-largemu}
          \end{figure}
          We see that for $\mu\gtrsim16$, this state is stable, which agrees with our theoretical
          expectation of spectral stability for large $\mu$
          (see also below). 
          
It is important to appreciate here that the linear instability
occurs practically immediately as we depart from the linear
limit

as the point spectrum eigenvalues collide with the
continuous spectrum leading to complex eigenvalue quartets
(via resonances) that persist for all $\mu \in [-8,0]$,
i.e., as the linear band is traversed. This type of 
{ instability} indeed persists for $\mu \lesssim 16$.

    \item \textbf{Defocusing case}

          {As in the focusing case, the existence results are
``exact'' given the equal and opposite amplitudes of 
solely the 6 excited adjacent sites within the lattice
which change as a function of $\mu$ (or $\varepsilon$),
modifying accordingly the mode power.
          For the stability of the states as a function of $\mu$, we can numerically track the magnitude of the eigenvalue of the linearization with the largest real part.
          This can be seen in Fig. \ref{fig:evalues-CLS-defocusing-rescaled-largemu}.}

          \begin{figure}[!tbp]
              \centering
              \includegraphics[width=.95\linewidth]{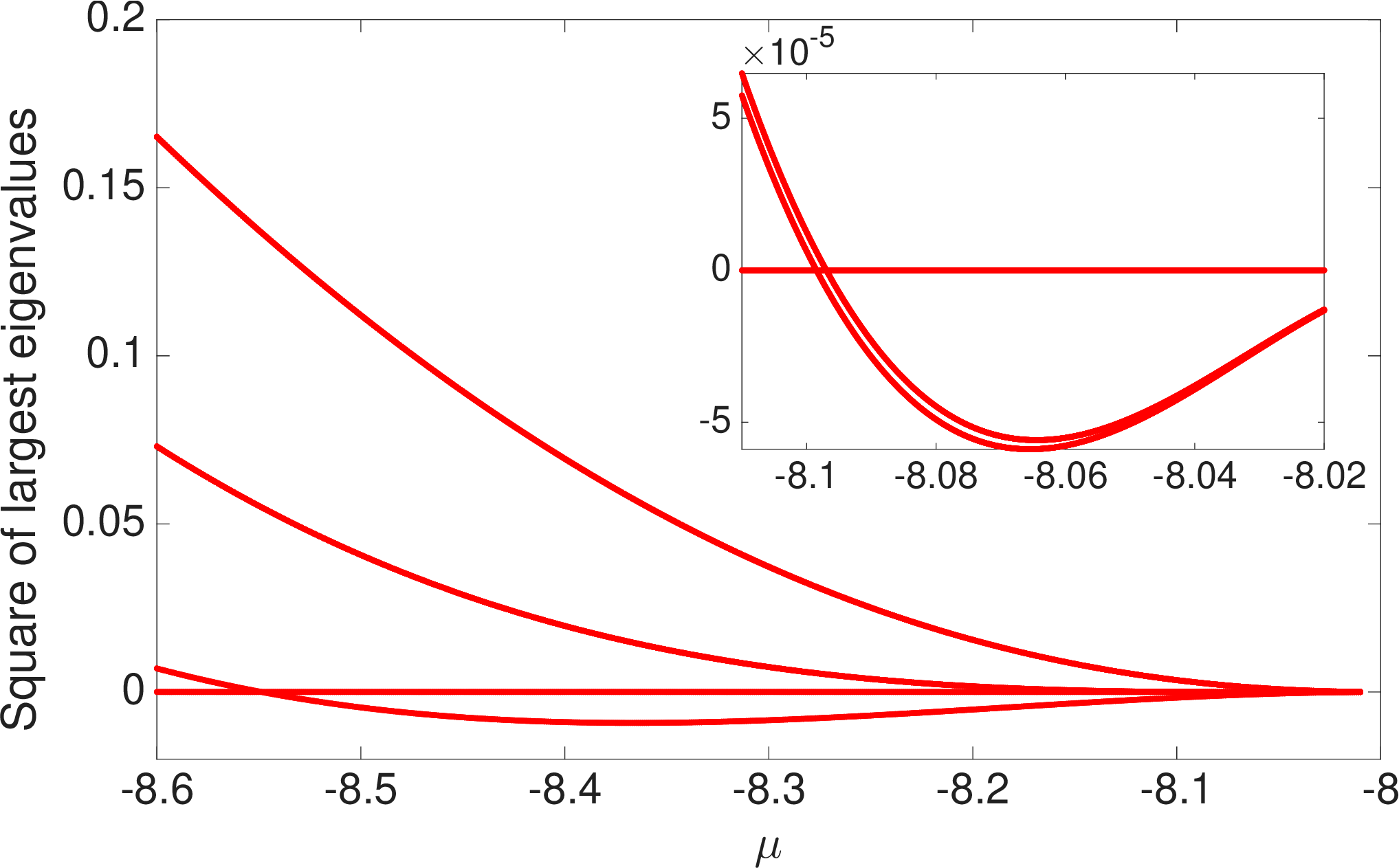}
              \caption{Square of the unstable eigenvalues of the linearization
              around the compactly supported state in the 
              defocusing case. { Inset:} Zoom around the first bifurcation point reveals the destabilization
              due { to} a pitchfork bifurcation at $\mu \approx
              -8.1$.}
              \label{fig:evalues-CLS-defocusing-rescaled-largemu}
          \end{figure}
          {We see that, as expected, the state is unstable for any value of $\mu<-8$
          (see also below).
It is important to appreciate here that the linear instability
occurs practically immediately as we depart from the linear
limit, and that there are two symmetry-breaking (i.e., pitchfork)
bifurcations, at $\mu\approx -8.1$ and $\mu\approx -8.55$. We can explore them in more detail to see the asymmetric states that arise from them.}
Notice that such bifurcations in the defocusing pyrochlore
lattice are directly reminiscent of corresponding
bifurcations in the Kagom{e} lattice, i.e., the 2d analog
of the pyrochlore setting considered herein~\cite{Vicencio13,shi2025stabilitytheoryflatband}.

{We can first look at the bifurcation at $\mu\approx-8.1$. As before, we can compute the square of the eigenvalues of the linearization around the new asymmetric state. We present 
the results in Fig. \ref{fig:evalues-CLS-defocusing-rescaled-largemu-asym1}.}
There we can see that the instability of the compactly supported
state is still present in the newly emergent state, yet the latter
is ``less unstable'' than the compactly supported one. Monitoring
the profile of this emergent state, we also
verify that it is
{\it asymmetric} in its nature.

\FloatBarrier
\begin{figure}[!htbp]
              \centering
              \includegraphics[width=1.0\linewidth]{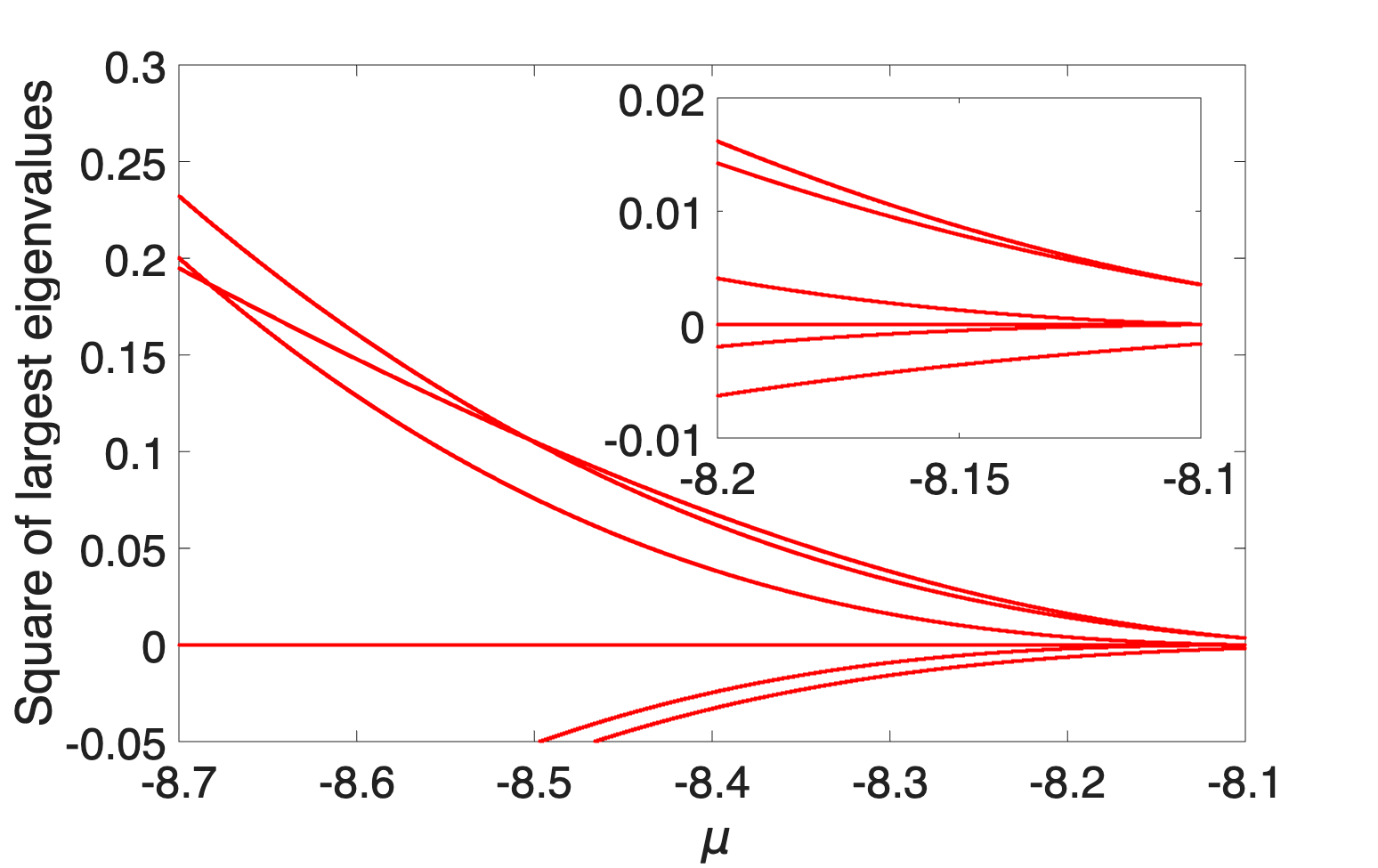}
              \caption{Square of the unstable eigenvalues of the linearization
              around the asymmetric state bifurcating from $\mu\approx-8.1$ in the 
              defocusing case. A zoom around the bifurcation point $\mu\approx-8.1$
              is shown in the inset.}
              \label{fig:evalues-CLS-defocusing-rescaled-largemu-asym1}
          \end{figure}

{Similarly, we can look at the bifurcation at $\mu\approx-8.55$. As before, we can compute the square of the eigenvalues of the linearization around the new asymmetric state. The associated stability results 
are shown in Fig. \ref{fig:evalues-CLS-defocusing-rescaled-largemu-asym2}.
There we can see that the instability of the compactly supported
state is still present in the newly emergent state, yet as in the previous case it is ``less unstable" than the compactly supported one. It is also worth noting that, as before, the resulting state is \emph{asymmetric} in its nature.}

\FloatBarrier
\begin{figure}[!htbp]
              \centering
              \includegraphics[width=.95\linewidth]{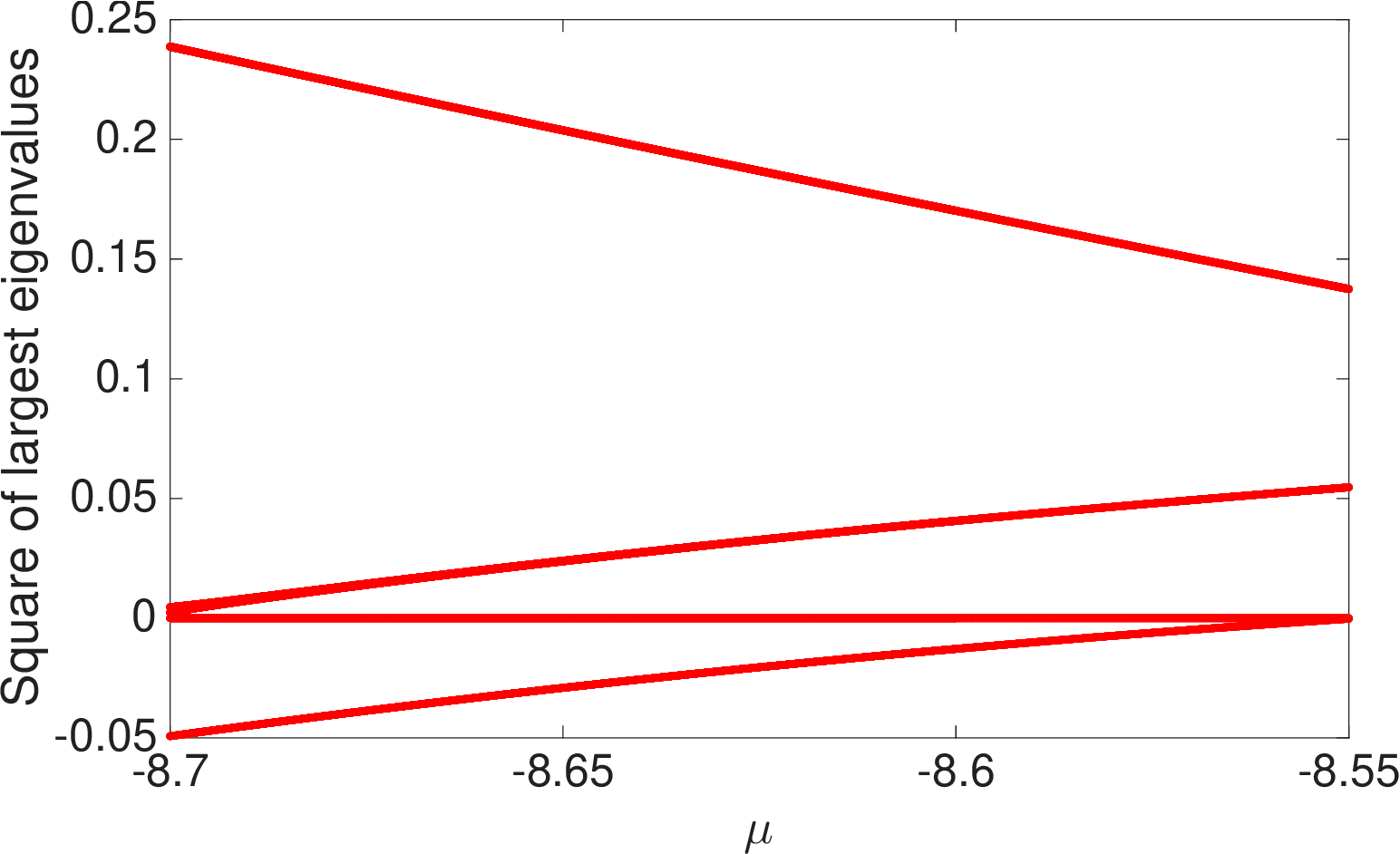}
              \caption{Square of the unstable eigenvalues of the linearization
              around the asymmetric state bifurcating from $\mu\approx-8.55$ in the 
              defocusing case.}
              \label{fig:evalues-CLS-defocusing-rescaled-largemu-asym2}
          \end{figure}

\end{itemize}

\subsubsection{Rescaled DNLS equation}
\begin{itemize}
    \item \textbf{Focusing case}

         We now explore the regime of large $\mu$, i.e.,
         small $\varepsilon$
in Fig. \ref{fig:evalues-CLS-focusing-rescaled-loglog}.
         \begin{figure}[!tbp]
              \centering
              \includegraphics[width=.95\linewidth]{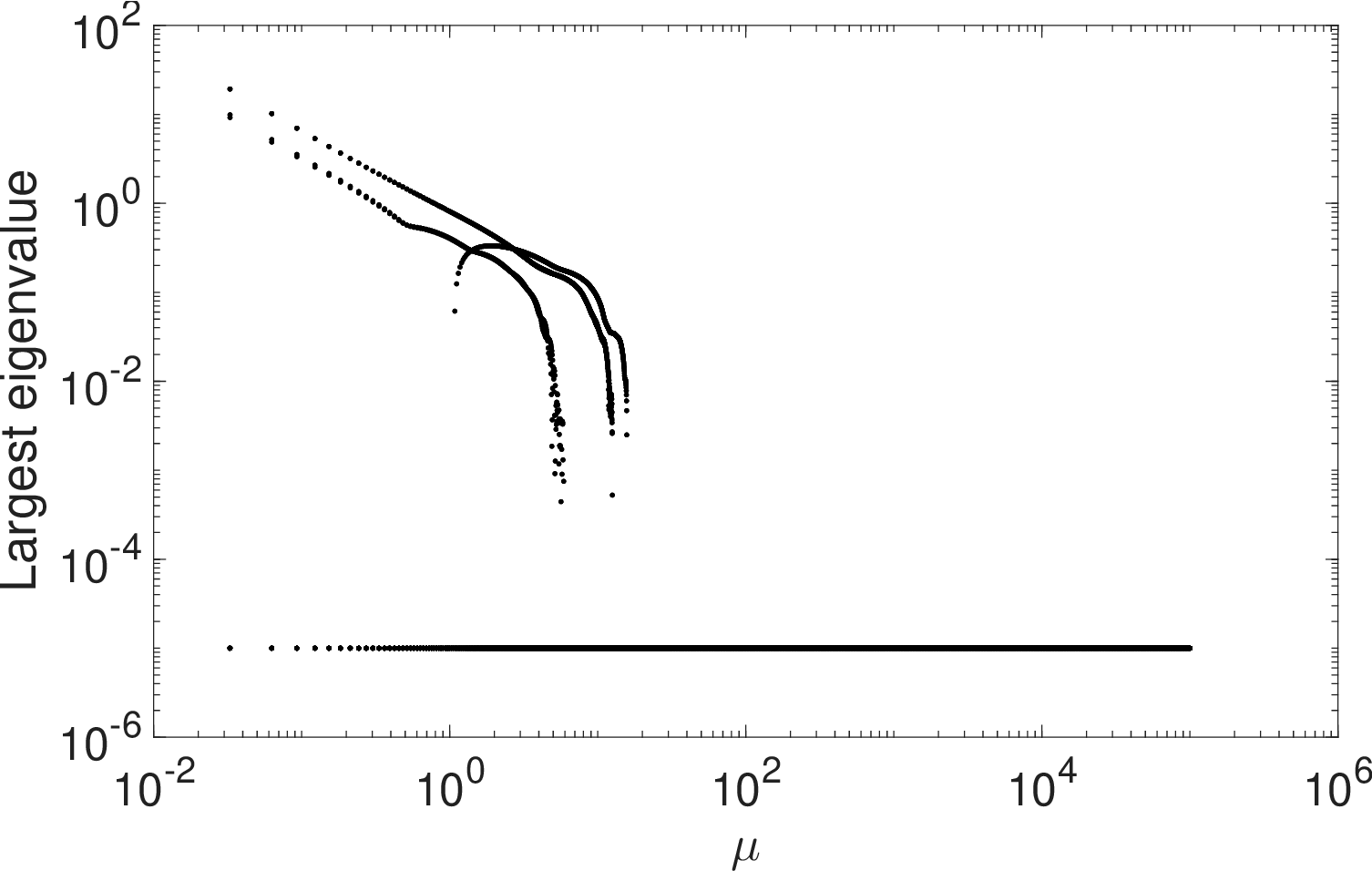}
              \caption{Positive eigenvalues of the linearization  around the compactly supported state in the focusing case for large $\mu$/small $\varepsilon$. Notice the
              { log-log} nature of the graph.}
              \label{fig:evalues-CLS-focusing-rescaled-loglog}
          \end{figure}
          We see that for $\mu\gtrsim16$, this state is stable, which agrees with our theoretical
          predictions for large $\mu$. 
          
          In addition to this, since for large $\mu$ the eigenvalues are purely imaginary,
          we may compare the results obtained numerically with the theoretical predictions
          for the growth of these imaginary eigenvalues. This can be observed in Fig. \ref{fig:evalues-CLS-focusing-rescaled-imaginary}.  As we can see, our predictions agree with the numerical results for large $\mu$, as
          expected from the AC limit. While the limiting behavior
          of such eigenvalues is very accurately captured,
          as $\mu$ becomes smaller/$\varepsilon$ becomes
          larger, the deviation is progressively larger, eventually
          leading to the collisions of eigenvalues with the
          continuous spectrum and instability for
          $\mu \lesssim 16$.

          \FloatBarrier
          \begin{figure}[!htbp]
              \centering
              \includegraphics[width=.95\linewidth]{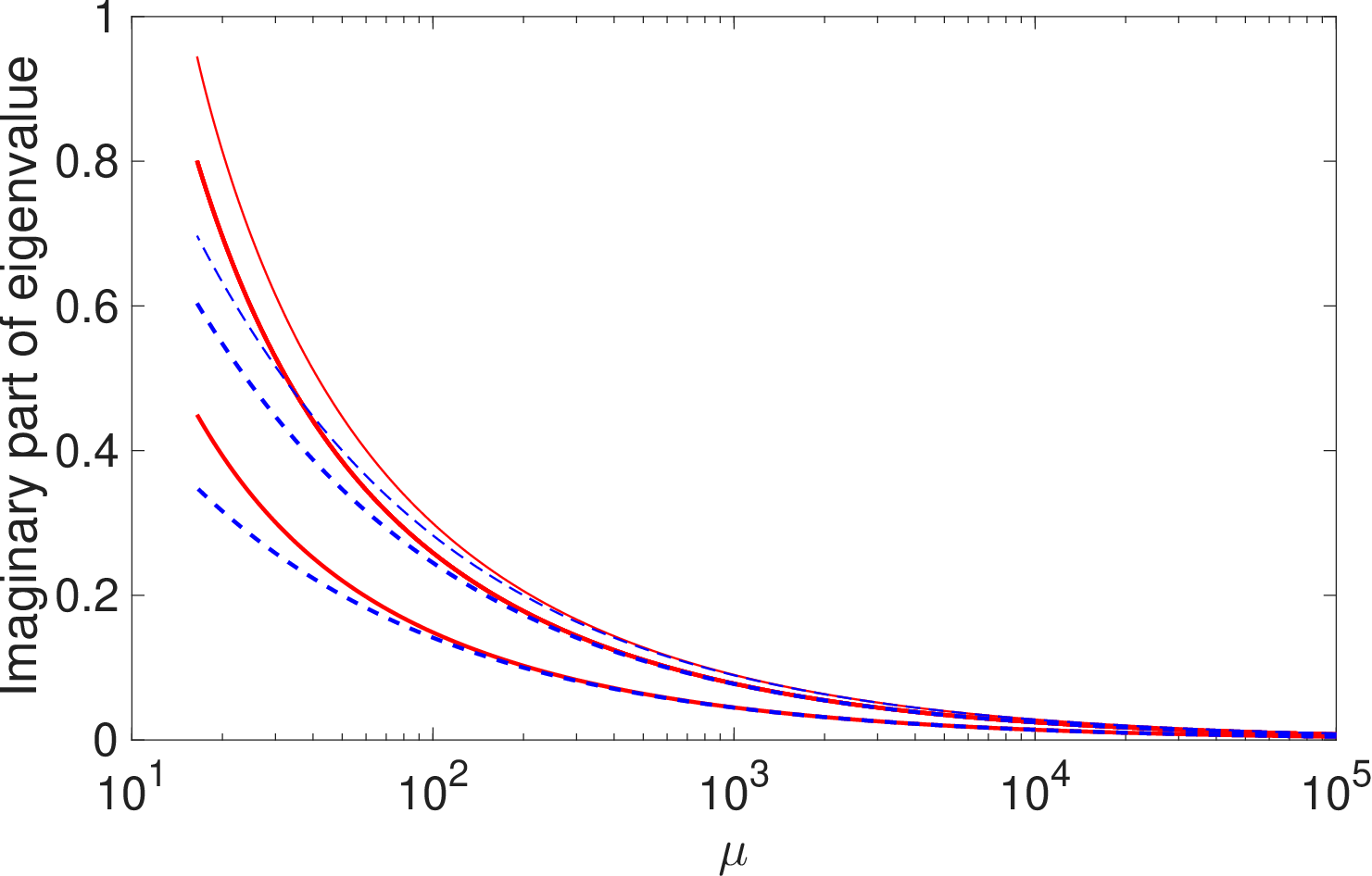}
              \caption{Imaginary part of the eigenvalues of the linearization approaching $i$ in the case of
              the compactly supported state in the focusing case. Dashed blue lines represent the theoretical predictions $i \sqrt{8\varepsilon}$, $i \sqrt{6\varepsilon}$, $i \sqrt{2\varepsilon}$; solid red lines represent the numerical results. Thicker lines represent double eigenvalues.}
              \label{fig:evalues-CLS-focusing-rescaled-imaginary}
          \end{figure}

    \item \textbf{Defocusing case}

           {We now explore the regime of large $\mu$, i.e.,
         small $\varepsilon$ for the defocusing setting.}

         {We find that, as expected, this state is unstable for any $\mu<-8$, i.e. in its whole range of existence. 
          
          In addition to this, we may compare the results obtained numerically with the theoretical predictions
          for the growth of these  eigenvalues. This can be observed in Fig. \ref{fig:evalues-CLS-defocusing-rescaled-real}.  Similar
          to the focusing case, our predictions agree with the numerical results for large $\mu$, as
          expected from the AC limit. 
          
          Once again, as $\mu$ becomes smaller/$\varepsilon$ becomes
          larger, the deviation is progressively more substantial, and in fact for small $\mu$ the eigenvalues seem to decrease towards 0, which is expected since this would represent convergence towards the spectrally stable vacuum state.}

          \FloatBarrier
          \begin{figure}[!htbp]
              \centering
              \includegraphics[width=.95\linewidth]{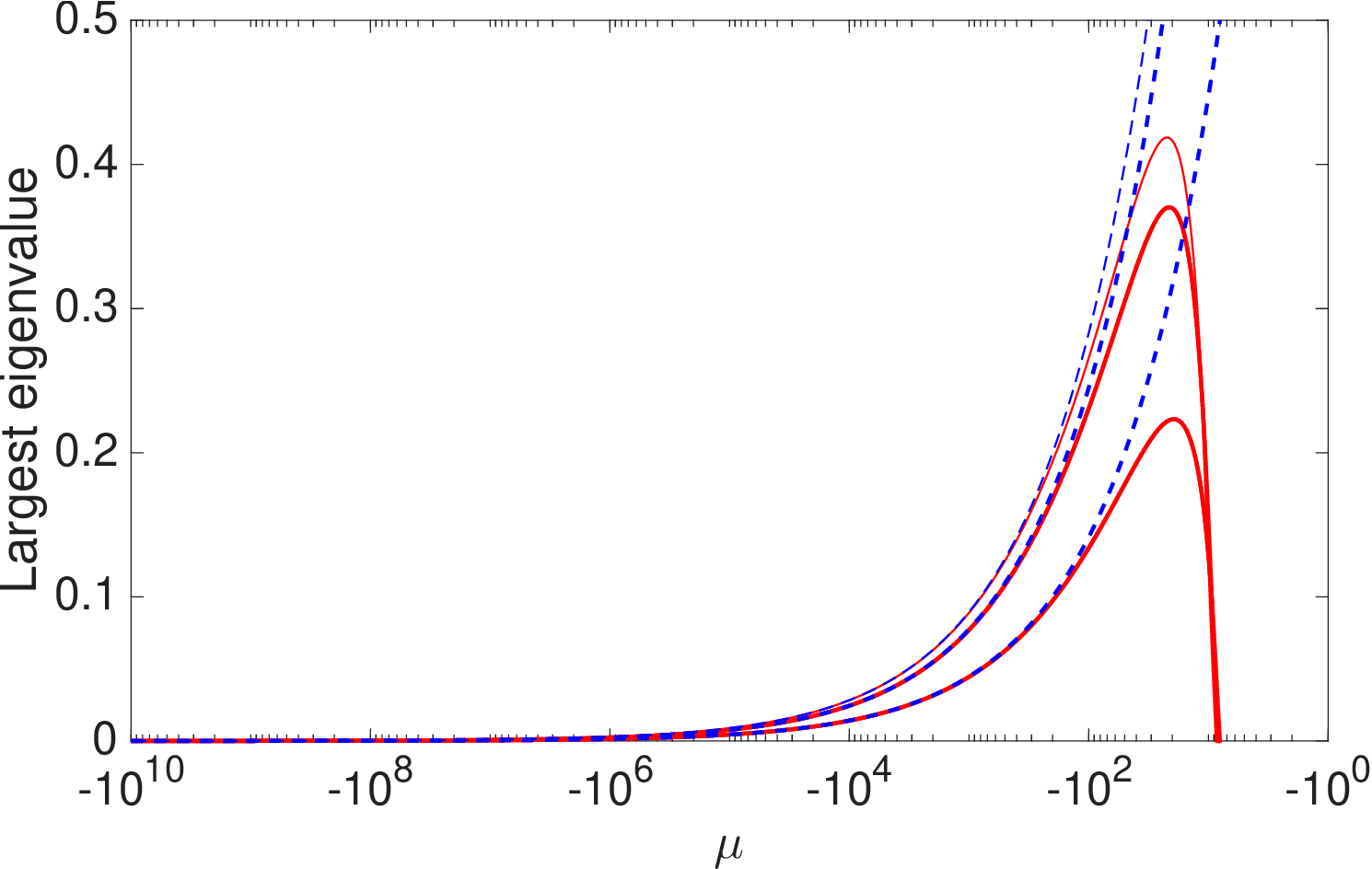}
              \caption{Largest eigenvalues of the linearization in the case of
              the compactly supported state of the defocusing
              case. Dashed blue lines represent the theoretical predictions $\sqrt{8\varepsilon}$, $\sqrt{6\varepsilon}$, $\sqrt{2\varepsilon}$; solid red lines represent the numerical results. Thicker lines represent double eigenvalues.}
              \label{fig:evalues-CLS-defocusing-rescaled-real}
          \end{figure}

\end{itemize}

\subsection{Vortices}
We will look at the results for both charge-1 and charge-2 vortices.
\subsubsection{Charge-1 vortex}
\FloatBarrier
\begin{figure}[!htbp]
 \centering
 \begin{subfigure}[b]{0.9\columnwidth}
     \centering
  \includegraphics[width=\linewidth]{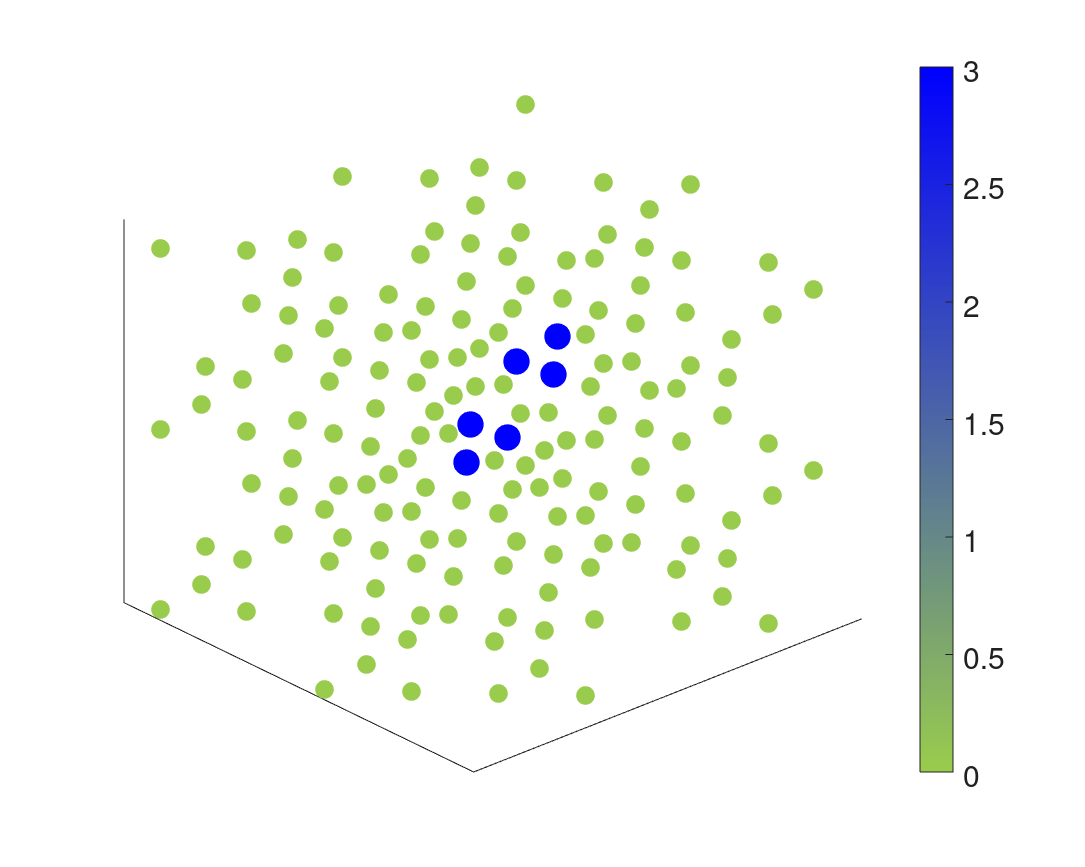}
  \end{subfigure}%
  \\
  \begin{subfigure}[b]{0.9\columnwidth}
    \centering
    \includegraphics[width=\linewidth]{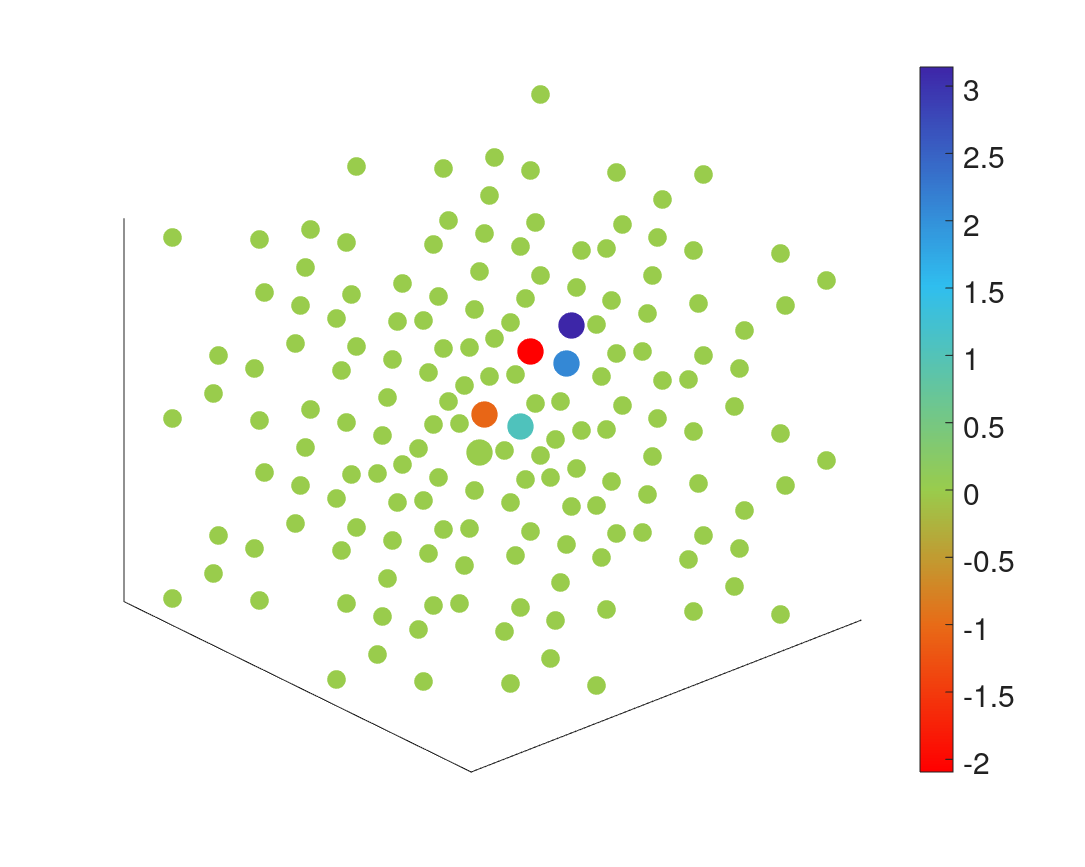}
  \end{subfigure}
 \caption{Charge-1 focusing vortex configuration. Density (top) and phase (bottom). Excited sites look bigger.}
 \label{fig:vortex}
\end{figure}
\begin{figure}[!htbp]
    \centering
    \includegraphics[width=.95\linewidth]{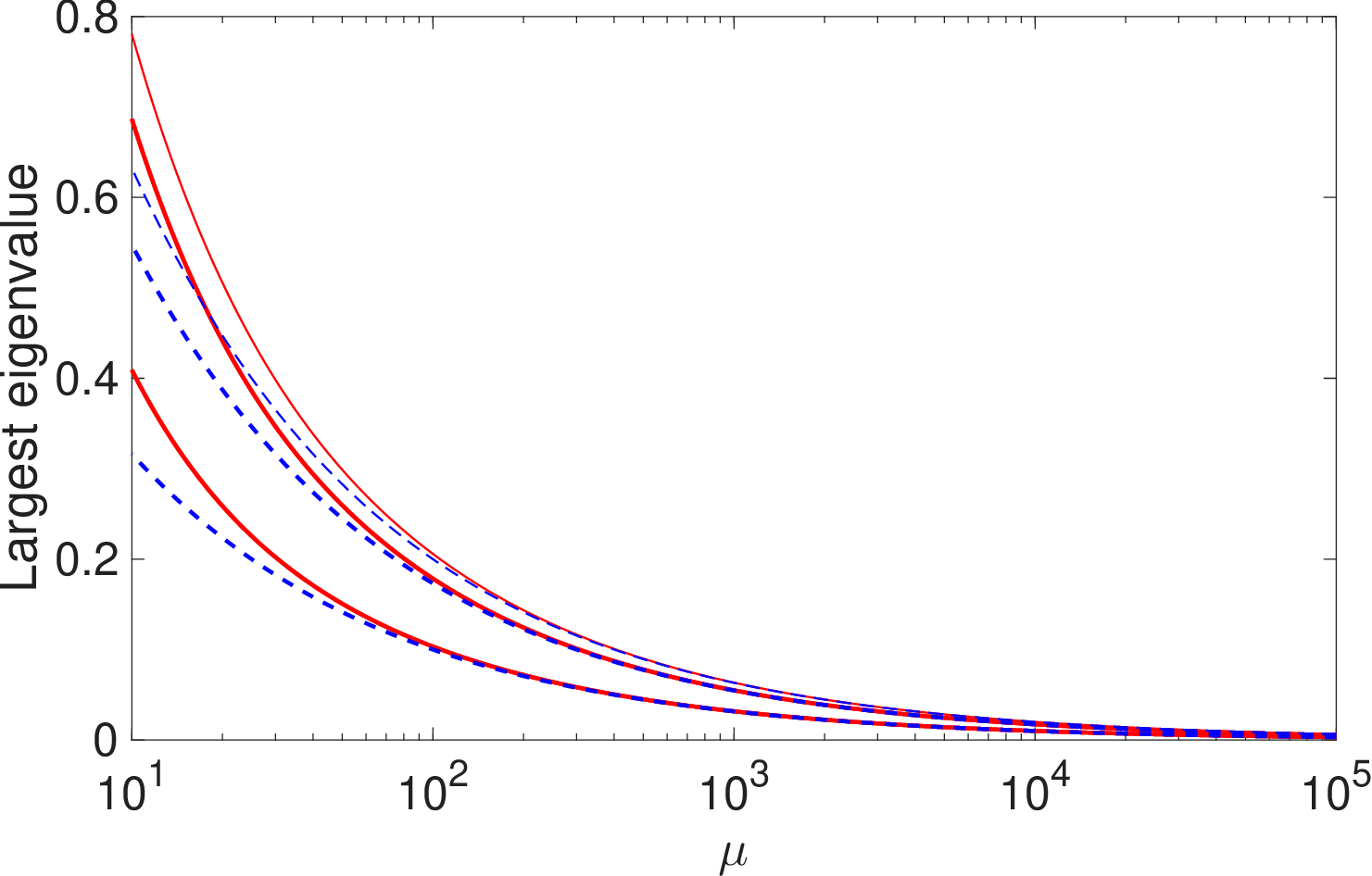}
    \caption{Largest eigenvalues of the linearization around
    a unit charge vortex in the focusing case. Dashed blue lines represent the theoretical predictions $\sqrt{4\varepsilon}$, $\sqrt{3\varepsilon}$, and $\sqrt{\varepsilon}$; solid red lines represent the numerical results. Thicker lines represent double eigenvalues.}
    \label{fig:evalues-vortex1-focusing-rescaled-theoretical}
\end{figure}

We start by considering a charge-1 vortex in a hexagonal configuration (cf. Fig.~\ref{fig:vortex}).
We now look at the corresponding nonlinear state, and continue it from the AC limit.
As in the previous cases, for the stability of the states as a function of $\mu$, we can numerically track the magnitude of the eigenvalues of the linearization with the largest real part. The values can be observed in Fig. \ref{fig:evalues-vortex1-focusing-rescaled-theoretical}.
As can be seen in the figure, the obtained results suggest that the state is unstable for $\mu>0$, which matches the instability expectation of the
theoretical prediction.
Additionally, we illustrate how  the real eigenvalues grow,
and compare this with our expectation based on the
anticontinuum limit analysis (cf. \eqref{eq:evalues-vortex1-theoretical}). 
The relevant comparison again showcases good agreement
in the vicinity of the AC limit, with the relevant deviation
becoming larger as we depart from that limit.

\FloatBarrier

\FloatBarrier
\subsubsection{Charge-2 vortex}
\begin{figure}[!htbp]
    \centering
    \begin{subfigure}[b]{0.9\columnwidth}
        \centering
        \includegraphics[width=\linewidth]{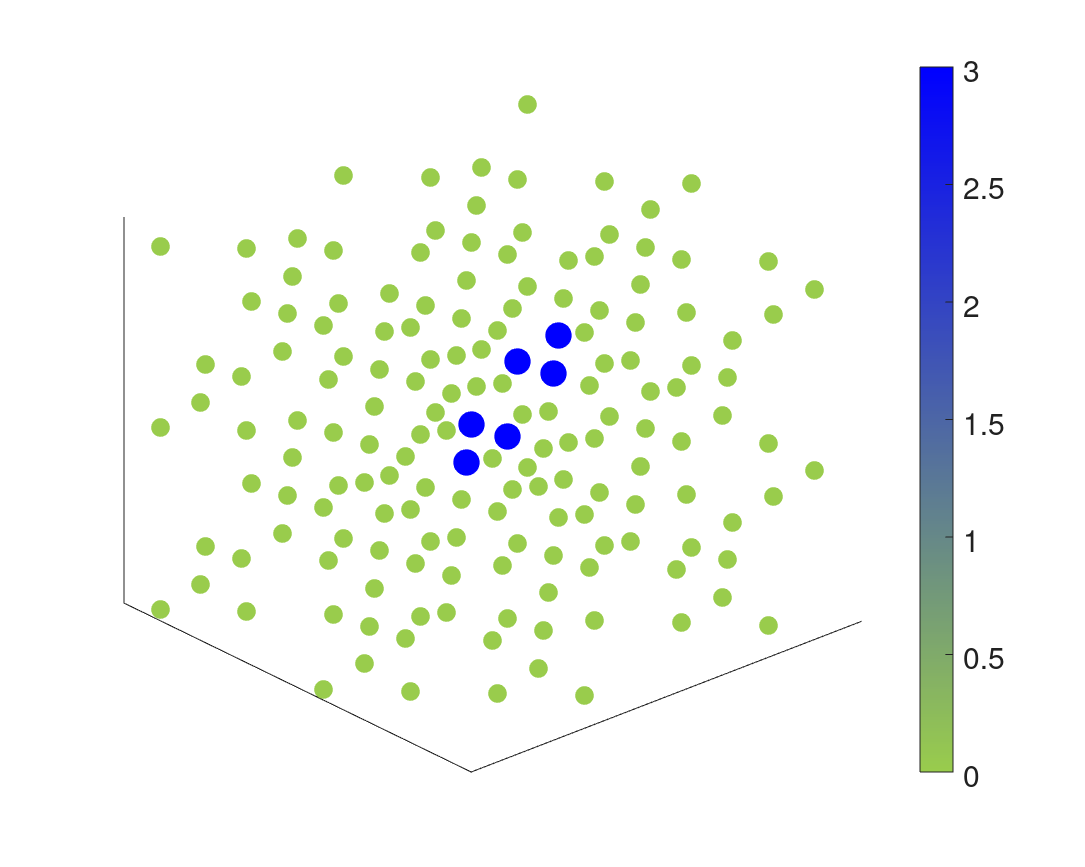}
    \end{subfigure}%
    \\
    \begin{subfigure}[b]{0.9\columnwidth}
        \centering
        \includegraphics[width=\linewidth]{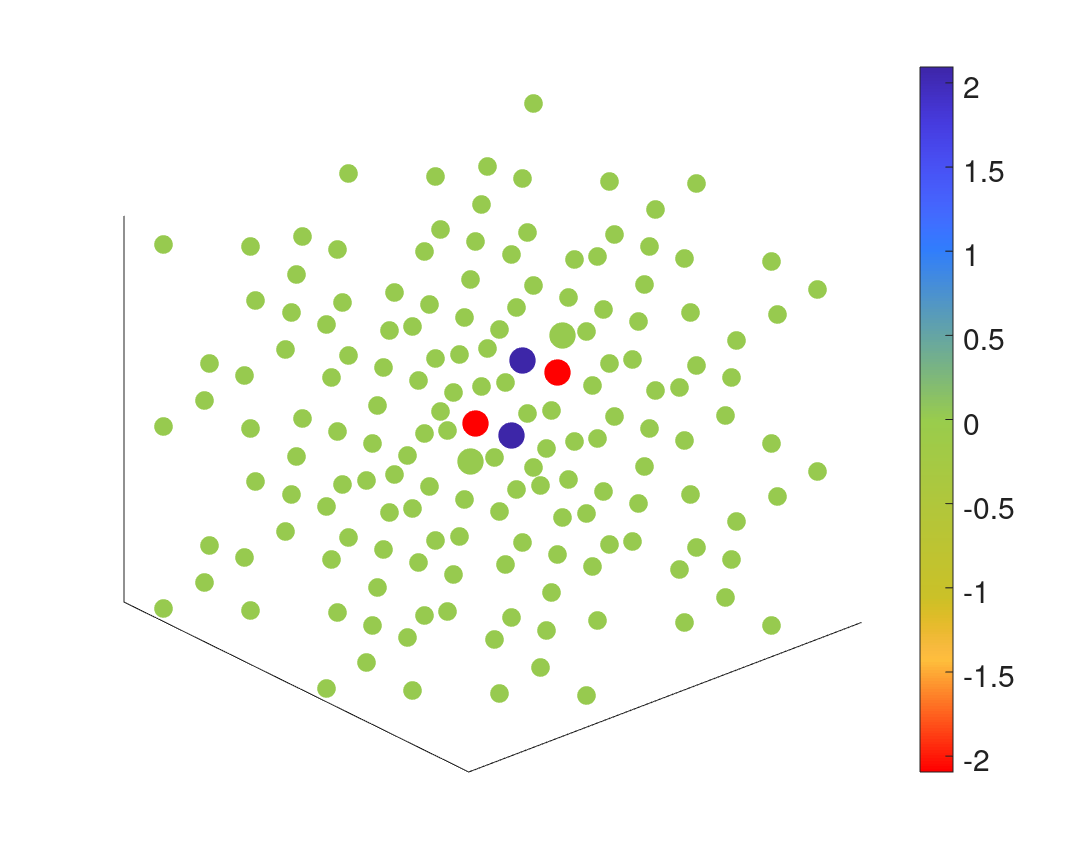}
    \end{subfigure}
    \caption{Charge-2 focusing vortex configuration. Density (top) and phase (bottom). Excited sites look bigger.}
    \label{fig:vortex2}
\end{figure}
\begin{figure}[!tbp]
    \centering
    \includegraphics[width=.95\linewidth]{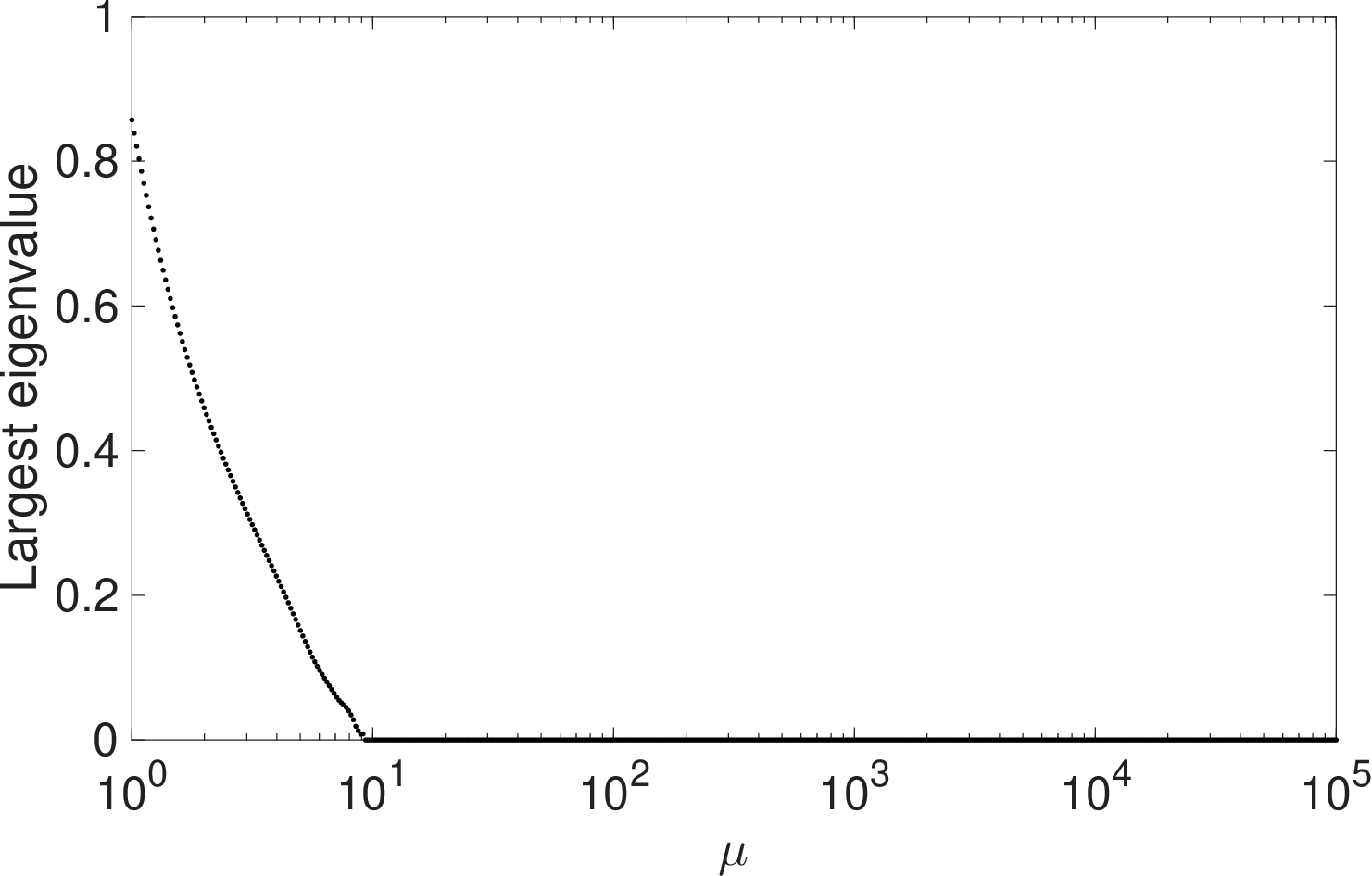}
    \caption{The largest eigenvalue of the linearization as a function of $\mu$ for the charge-2 vortex in the focusing case.}
    \label{fig:evalues-vortex2-focusing-rescaled}
\end{figure}
\begin{figure}[!tbp]
    \centering
    \includegraphics[width=.95\linewidth]{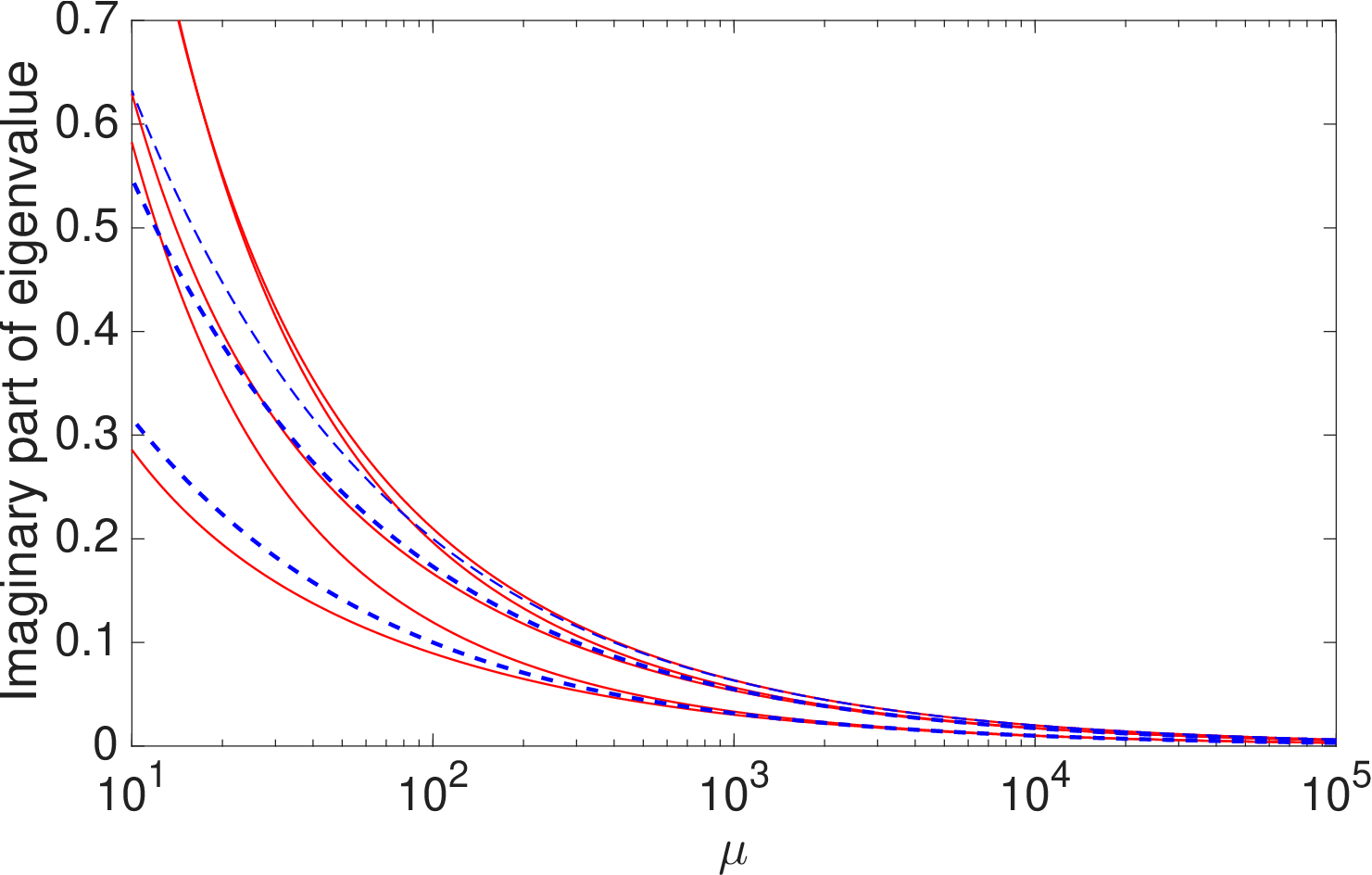}
    \caption{Imaginary parts of the eigenvalues of the linearization approaching $i$ around a charge-2 vortex in the focusing case. Dashed blue lines represent the theoretical predictions $\sqrt{4\varepsilon}$, $\sqrt{3\varepsilon}$, and $\sqrt{\varepsilon}$; solid red lines represent the numerical results. Thicker lines represent double eigenvalues.}
    \label{fig:evalues-vortex2-focusing-rescaled-theoretical}
\end{figure}

We now consider a charge-2 vortex in a hexagonal configuration (i.e., the phase difference between neighboring nodes is now $2\pi/3$), see Fig. \ref{fig:vortex2}.
As in the previous cases, for the stability of the states as a function of $\mu$, we can numerically track the magnitude of the eigenvalue of the linearization with the largest real part. The values can be observed in Fig. \ref{fig:evalues-vortex2-focusing-rescaled}.
As we can discern in Fig.~\ref{fig:evalues-vortex2-focusing-rescaled}, the results obtained suggest that the state is stable for $\mu\gtrsim10$, in line with what is expected on the basis of
our AC limit analysis.
Additionally, we may look at how the imaginary eigenvalues starting at zero grow near the AC limit of $\varepsilon=0$,
and compare this with our previous theoretical predictions. This can be observed in Fig. \ref{fig:evalues-vortex2-focusing-rescaled-theoretical}. Here, too, the purely imaginary 
predicted eigenvalue pairs (cf. \eqref{eq:evalues-vortex2-theoretical}) seem
to be accurately captured in the numerical computations.
It is only when the first of these pairs collides with 
the continuous spectrum that the relevant instability develops.
This kind of agreement is also identified in the defocusing
case for the vortices of charge $S=1$ and $S=2$, although
as noted above their respective stability properties are
reversed there. These results are omitted here for brevity.

\section{Conclusions and Future Work}

In the present work we have investigated a discrete nonlinear topological dynamical lattice model in three spatial dimensions.  Specifically, we have studied the DNLS equation on a pyrochlore lattice.  The latter arises in many physical contexts including frustrated magnetism.  It harbors flat bands and is a 3d generalization of the 2d Kagom{e} lattice, bearing a number of associated stationary states.  We have studied steady state and dynamical properties of this system and carried out a systematic analysis of the associated nonlinear equilibria.  In particular, both discrete solitons as well as (charge-1 and charge-2) vortices were numerically studied in the context of their existence and stability.  Notably, the pyrochlore lattice was found to possess compactly supported nonlinear eigenstates that arise from the flat bands of the linear spectrum.  In the focusing nonlinearity case these modes possess oscillatory instabilities in a certain range of propagation constants.  On the other hand, in the defocusing case, such waveforms are found
to be generically unstable. The vortical structures bear similar features to the
Kagom{e} setting with charge $2$ vortices being more stable
than unit-charge ones in the focusing nonlinearity realm, while
the reverse is true in the defocusing case. 

Our findings open new avenues for investigating the intriguing interplay of topology and flat bands in three dimensional dispersive nonlinear lattices. On the one hand, it would be useful to gain further insight of the dynamics of this system. To this effect, it would be interesting to extend this study to the dynamical properties of the studied solitons or vortices, including potentially their collisions. 
Moreover, while here we have focused predominantly on 
quasi-two-dimensional configurations to ensure that they can
survive in 3d while potentially maintaining their robustness therein, studying genuinely 3d structures in suitable
contours (similar, e.g., to what was done for cubic
lattices in~\cite{LUKAS2008339}) is of particular interest
in its own right.
Moreover, the methods employed in this paper should in principle be applicable to the study of other nonlinear dispersive lattice models, such as the Chern-Hopf insulator \cite{Dutta2024}. Like the pyrochlore lattice, they exhibit 3d flat bands with non-trivial topology and would thus be of great interest for future investigations.
Such studies are currently in progress and will be reported
in future publications.

\begin{acknowledgments}
This material is based upon work supported by the National Science
Foundation under Grant No. DMS-1928930, while F.P.R. was in
residence at the Simons Laufer Mathematical Sciences Institute in
Berkeley, California, during the Fall 2025 semester. The work at 
LANL was carried out under the auspices of the US Department of 
Energy NNSA under Contract No. 89233218CNA000001.
This research was supported by the U.S. National Science Foundation under the award PHY-2408988 (P.G.K). 
This research was partly conducted while P.G.K. was  visiting the Okinawa Institute of Science and
Technology (OIST) through the Theoretical Sciences Visiting Program (TSVP), the University of
Sydney through the visitor program of the Sydney Mathematical Research Institute (SMRI) and the Department of Mechanical Engineering at Seoul National 
University through a Fulbright Fellowship. Their support is gratefully acknowledged.
Finally, this work was also  supported by a grant from the Simons Foundation [SFI-MPS-SFM-00011048, P.G.K]. 
\end{acknowledgments}

\bibliography{apssamp_v2}

\end{document}